\documentclass{article}
\usepackage{amssymb}

\begin{document}


\title{The Dirac Equation Near Centenary: a Contemporary Introduction to the Dirac Equation Consideration}

\author{V.M. Simulik\\
Institute of Electron Physics of the NAS of Ukraine\\
21 Universitetska Str. Uzhgorod 88017, Ukraine\\
Email: vsimulik@gmail.com}
\date{11. 09. 2024}

\maketitle

\begin{abstract} More then 35 approaches to the Dirac equation derivation are presented. The various physical principles and mathematical methods are used. A review of well-known and not enough known contributions to the problem is given, the unexpected and unconventional derivations are presented as well. Three original approaches to the problem suggested by the author are considered as well. They are (i) the generalization of H. Sallhofer derivation, (ii) the obtaining of the massless Dirac equation from the Maxwell equations in maximally symmetrical form, (iii) the derivation of the Dirac equation with nonzero mass from the relativistic canonical quantum mechanics of the fermion-antifermion spin s=1/2 doublet. Today we are able to demonstrate new features of our derivations given in original papers. In some sense the important role of the Dirac equation in contemporary theoretical physics is demonstrated. A criterion for the usefulness of one or another derivation of the Dirac equation has been established.
\end{abstract}


\pagestyle{plain}
\pagenumbering{arabic}
\setcounter{page}{1}

\section{Contents}

\begin{flushleft}
2. Introduction \hspace{1cm}\hspace{1cm}\hspace{1cm}\hspace{1cm}\hspace{1cm}\hspace{1cm}\hspace{1cm}\hspace{1cm}\hspace{1cm}\hspace{1cm}\hspace{0.2cm}  2
3. Variants of the Dirac equation derivation \hspace{1cm}\hspace{1cm}\hspace{1cm}\hspace{1cm}\hspace{1cm}\hspace{0.8cm} 3

\hspace{0.3cm} Derivation of the Dirac equation from start \hspace{1cm}\hspace{1cm}\hspace{1cm}\hspace{1cm}\hspace{1cm}\hspace{0.5cm} 3

\hspace{0.5cm} 1. Dirac's derivation \hspace{1cm}\hspace{1cm}\hspace{1cm}\hspace{1cm}\hspace{1cm}\hspace{0.8cm}\hspace{0.8cm}\hspace{0.8cm}\hspace{0.8cm}\hspace{0.6cm} 3

\hspace{0.5cm} 2. Van der Waerden--Sakurai derivation \hspace{1cm}\hspace{1cm}\hspace{1cm}\hspace{1cm}\hspace{1cm}\hspace{0.9cm} 4

\hspace{0.5cm} 3. Briefly on "factorization" of the Klein--Gordon operator \hspace{1cm}\hspace{1cm}\hspace{1cm} 4

\hspace{0.5cm} 4. Principle of least action for the spinor field \hspace{1cm}\hspace{1cm}\hspace{1cm}\hspace{1cm}\hspace{1cm} 5

\hspace{0.5cm} 5. Ryder's derivation \hspace{1cm}\hspace{1cm}\hspace{1cm}\hspace{1cm}\hspace{1cm}\hspace{0.7cm}\hspace{1cm}\hspace{1cm}\hspace{1cm} 5

\hspace{0.5cm} 6. Start from the initial geometric properties of the space-time and electron \hspace{0.4cm} 5

\hspace{0.5cm} 7. Origin in Bargman--Wigner classification \hspace{1cm}\hspace{1cm}\hspace{1cm}\hspace{1cm}\hspace{1cm}\hspace{0.3cm} 5

\hspace{0.5cm} 8. The partial case of the Bargman--Wigner equation \hspace{1cm}\hspace{1cm}\hspace{1cm}\hspace{0.8cm} 5

\hspace{0.5cm} 9. Inverse Foldy--Wouthuysen transformation \hspace{1cm}\hspace{1cm}\hspace{1cm}\hspace{1cm}\hspace{0.1cm}\hspace{1cm} 5

\hspace{0.5cm} 10. Feynman's derivation \hspace{1cm}\hspace{1cm}\hspace{1cm}\hspace{1cm}\hspace{1cm}\hspace{0.7cm}\hspace{1cm}\hspace{1cm}\hspace{0.4cm} 6

\hspace{0.3cm} Unexpected derivations \hspace{1cm}\hspace{1cm}\hspace{1cm}\hspace{1cm}\hspace{1cm}\hspace{0.5cm}\hspace{1cm}\hspace{1cm}\hspace{1cm}\hspace{0.1cm} 6

\hspace{0.5cm} 11. Sallhofer's derivation from the Maxwell electrodynamics in medium \hspace{1cm} 6

\hspace{0.5cm} 12. In addition to Ryder's derivation \hspace{1cm}\hspace{1cm}\hspace{1cm}\hspace{1cm}\hspace{1cm}\hspace{1cm}\hspace{0.3cm} 6

\hspace{0.5cm} 13. Derivation from the Langevin equation for a two-valued process \hspace{1cm}\hspace{0.6cm} 6

\hspace{0.5cm} 14. Derivation from the conservation law of spin 1/2 current \hspace{1cm}\hspace{1cm}\hspace{0.7cm} 7

\hspace{0.5cm} 15. Derivation from the master equation of Poisson process \hspace{1cm}\hspace{1cm}\hspace{0.9cm} 7

\hspace{0.5cm} 16. Derivation from the relativistic Newton's second law \hspace{1cm}\hspace{1cm}\hspace{1cm}\hspace{0.3cm} 7

\hspace{0.5cm} 17. A geometrical derivation of the Dirac equation \hspace{1cm}\hspace{1cm}\hspace{1cm}\hspace{1cm}\hspace{0.2cm} 8

\hspace{0.5cm} 18. Start from the geodesic equation \hspace{1cm}\hspace{1cm}\hspace{1cm}\hspace{1cm}\hspace{1cm}\hspace{1cm}\hspace{0.3cm} 8

\hspace{0.5cm} 19. Derivation from a generally covariant field equation for gravitation and electromagnetism \hspace{1cm}\hspace{1cm}\hspace{1cm}\hspace{1cm}\hspace{1cm}\hspace{1cm}\hspace{1cm}\hspace{1cm}\hspace{1cm}\hspace{0.9cm} 8

\hspace{0.5cm} 20. Derivation from the unique conditions for wave function \hspace{1cm}\hspace{1cm}\hspace{0.8cm} 8

\hspace{0.5cm} 21. Derivation by factorizing the algebraic relation satisfied by the classical Hamiltonian \hspace{1cm}\hspace{1cm}\hspace{1cm}\hspace{1cm}\hspace{1cm}\hspace{1cm}\hspace{1cm}\hspace{1cm}\hspace{1cm}\hspace{1cm}\hspace{0.7cm} 8

\hspace{0.5cm} 22. Origin in conformal differential geometry \hspace{1cm}\hspace{1cm}\hspace{1cm}\hspace{1cm}\hspace{1cm}\hspace{0.1cm} 9

\hspace{0.5cm} 23. Derivation from principles of information processing \hspace{1cm}\hspace{1cm}\hspace{1cm}\hspace{0.4cm} 9

\hspace{0.5cm} 24.  A method to derive the anticommutation properties of the Dirac matrices without relying on the squaring of the Dirac Hamiltonian \hspace{1cm}\hspace{1cm}\hspace{1cm}\hspace{0.8cm} 9

\hspace{0.5cm} 25. Equation of motion for the spherical spin model \hspace{1cm}\hspace{1cm}\hspace{1cm}\hspace{1cm}\hspace{0.1cm} 9

\hspace{0.5cm} 26. Dirac equation from the strands and tangles \hspace{1cm}\hspace{1cm}\hspace{1cm}\hspace{1cm}\hspace{0.6cm} 9

\hspace{0.5cm} 27. A diffusion model for the Dirac equation \hspace{1cm}\hspace{1cm}\hspace{1cm}\hspace{1cm}\hspace{0.9cm}\hspace{0.1cm} 10

\hspace{0.5cm} 28. A variational principle for the Feynman's derivation \hspace{1cm}\hspace{1cm}\hspace{1cm}\hspace{0.3cm} 10

\hspace{0.5cm} 29. The Dirac equation from scratch \hspace{1cm}\hspace{1cm}\hspace{1cm}\hspace{1cm}\hspace{1cm}\hspace{1cm}\hspace{0.2cm} 10

\hspace{0.5cm} 30. Another attempt from scratch \hspace{1cm}\hspace{1cm}\hspace{1cm}\hspace{1cm}\hspace{1cm}\hspace{1cm}\hspace{0.6cm} 10

\hspace{0.5cm} 31. Scale-relativistic derivations of Pauli and Dirac equations \hspace{1cm}\hspace{1cm}\hspace{0.4cm} 11

\hspace{0.5cm} 32. Recent derivation of Dirac equation from the stochastic optimal control principles of quantum mechanics \hspace{1cm}\hspace{1cm}\hspace{0.4cm}\hspace{1cm}\hspace{1cm}\hspace{1cm}\hspace{1cm}\hspace{1cm} 11

\hspace{0.3cm} Our approaches to the problem of the Dirac equation derivation \hspace{1cm}\hspace{1cm}\hspace{0.2cm} 11

\hspace{0.5cm} 33. Generalization of Sallhofer's derivation \hspace{1cm}\hspace{1cm}\hspace{1cm}\hspace{1cm}\hspace{1cm}\hspace{0.2cm} 12

\hspace{0.5cm} 34. Origin in maximally symmetrical form of the Maxwell equations \hspace{1cm}\hspace{0.3cm} 14

\hspace{0.5cm} 35. Derivation of the Dirac equation from the relativistic canonical quantum mechanics \hspace{1cm}\hspace{1cm}\hspace{1cm}\hspace{0.4cm}\hspace{1cm}\hspace{1cm}\hspace{1cm}\hspace{1cm}\hspace{0.4cm}\hspace{1cm}\hspace{1cm}\hspace{1cm} 18

\hspace{0.3cm} Derivation of the Dirac-like equations \hspace{1cm}\hspace{1cm}\hspace{1cm}\hspace{1cm}\hspace{1cm}\hspace{1cm}\hspace{0.2cm} 21

\hspace{0.5cm} 36. On the derivation of the Dirac equation in N and six spatial dimensions \hspace{0.2cm} 21

\hspace{0.5cm} 37. Derivation of the fractional Dirac equation \hspace{1cm}\hspace{1cm}\hspace{1cm}\hspace{1cm}\hspace{0.7cm} 21

\hspace{0.5cm} 38. On the derivation of the Dirac-like equations for spin 3/2, 2 and arbitrary spin \hspace{1cm}\hspace{1cm}\hspace{1cm}\hspace{1cm}\hspace{1cm}\hspace{1cm}\hspace{1cm}\hspace{1cm}\hspace{1cm}\hspace{1cm}\hspace{1cm}\hspace{0.8cm} 22

4. Conclusions and perspectives \hspace{1cm}\hspace{1cm}\hspace{1cm}\hspace{1cm}\hspace{1cm}\hspace{1cm}\hspace{1cm}\hspace{0.5cm} 22

References \hspace{1cm}\hspace{1cm}\hspace{1cm}\hspace{1cm}\hspace{1cm}\hspace{1cm}\hspace{1cm}\hspace{1cm}\hspace{1cm}\hspace{1cm}\hspace{0.8cm} 23

\end{flushleft}

\section{Introduction}
\setcounter{page}{2}
Started more than 10 years ago from the natural idea to determine the place of our original derivation of the Dirac equation among the near centenary long suggestions of other authors we are able today to present the contemporary interesting general picture.

The Dirac equation [1] is one of the fundamental equations of modern theoretical physics. It is in service near 100 years (1928--2024). The application today is much wider than the areas of quantum mechanics [2, 3], quantum field theory, atomic and nuclear physics, solid state physics. Let us recall only that the first analysis of this equation enabled Dirac to give the theoretical prediction of the positron, which was discovered experimentally by Anderson in 1932. The recent well-known application [4] of the massless Dirac equation to the graphene ribbons is an example of the contemporary possibilities of this equation. The Dirac equation inspired the development of such mathematical models as the matrix representations of the Clifford algebras [5, 6] and corresponding groups [7, 8] with the wide-range applications in the theory of high energy physics.

The successful derivation of some equation of mathematical physics is the first step to successful application. In such process the essence of the corresponding model of nature, the mathematical principles and the physical foundations are visualized. Here we deal with the different approaches to the problem of the Dirac equation derivation

In this communication together with some review the original investigation of the problem of the Dirac equation derivation is presented. The different approaches, which are based on the various mathematical and physical principles, are considered (much more than 35 methods). Meanwhile, the importance of place of the Dirac equation in modern theoretical physics is discussed.

In recent publications [9--16] we were able to extend the domain of the Dirac equation application. We proved [9--16] that this equation
has not only fermionic but also the bosonic features, can describe not only fermionic, but also the bosonic states.

Note that in modern quantum field theory the Dirac equation is applied not only to the spin 1/2 spinor field description but in its different generalizations is the base for the construction of the models of higher and arbitrary spin elementary particles as well.

In our approach to quantum field theory [3] the Dirac equation is derived  from the 4-component formalism of relativistic canonical quantum mechanics (RCQM). In order to determine exactly the place of our derivation [15--16] among the other known methods we consider below the different ways of the Dirac equation derivation.

Thus, it is evident that the methods of the derivation of the Dirac equation cause interest of the researchers. Indeed, the new ways of the Dirac equation derivation automatically visualize the ground principles, which are in the foundations of the modern description of the elementary particles. Hence, the active consideration of the different ways of the Dirac equation derivation is the subject of many contemporary publications. Usually the start from the different basic principles and assumptions is considered.

An introduction to the Dirac equation consideration is based on different variants of its derivation. Below a review of 36 different ways of the Dirac equation derivation is given.

Here the system of units  $\hbar=c=1$ is chosen, the metric tensor in Minkowski space-time $\mathrm{M}(1,3)$ is given by $g^{\mu\nu}=g_{\mu\nu}=g^{\mu}_{\nu}, \, \left(g^{\mu}_{\nu}\right)=\mathrm{diag}\left(1,-1,-1,-1\right), \, x_{\mu}=g_{\mu\nu}x^{\mu},$ and summation over the twice repeated indices is implied. Note that in quantum-mechanical theories, where the time variable plays the role of a parameter, the rigged Hilbert space $\mathrm{S}^{3,4}\subset\mathrm{H}^{3,4}\subset\mathrm{S}^{3,4*}$ is used. Nevertheless, in the formalism of the manifestly covariant field theory the rigged Hilbert space is taken as $\mathrm{S}^{4,4}\subset\mathrm{H}^{4,4}\subset\mathrm{S}^{4,4*}$ ($\mathrm{S}^{4,4}$ is the corresponding Schwartz test function space).

\section{Variants of the Dirac equation derivation}

\begin{center}
\textbf{Derivation of the Dirac equation from start}
\end{center}

\textbf{1. Dirac's derivation.} At first, one should note \textit{the elegant derivation given by Paul Dirac} in his book [17] (of course, this consideration is based on origins of [1]). Until today it is very interesting for the readers to feel Dirac's way of thinking and to follow his logical steps. One can follow his start from the principles of linearity, manifestly Lorentz invariance, relationship to the Klein--Gordon equation and finish in determination of $\alpha$ and $\beta$ matrices explicit forms.

Nevertheless, the Dirac's consideration of the Schr$\mathrm{\ddot{o}}$dinger--Foldy equation (our suggestion to call this equation as the Sch\"odinger--Foldy equation is considered after few steps below)
$$
\left\{p_{0}-\left(m^{2}c^{2}+p^{2}_{1}+p^{2}_{2}+p^{2}_{3}\right)^{\frac{1}{2}}\right\}\psi=0,
$$
which was essentially used in his derivation [17], was quite hasty. Especially his assertion that Schr$\mathrm{\ddot{o}}$dinger--Foldy equation is unsatisfactory from the point of view of the relativistic theory. Dirac's doubts were overcome later after the Foldy--Wouthuysen (FW)  consideration in the papers [18--20], where the corresponding canonical representation of the Dirac equation has been suggested and interpreted.

Today the significance of the Sch\"odinger--Foldy equation is confirmed by few hundreds publications about FW (see, e. g., [21--37] and the references therein) and the spinless Salpeter [38--55] equations (together with its generalizations), which have wide-range application in contemporary theoretical physics. Our contribution in consideration of the FW and Schr$\mathrm{\ddot{o}}$dinger--Foldy equations is presented in [56--58] and [3, 15, 16]. Reviewing the papers [21--37] we tried to demonstrate the generalisations of the FW transformations for an arbitrary spin.

Suggestion to call the main equation of the contemporary $N$-component relativistic canonical quantum mechanics as the Schr$\mathrm{\ddot{o}}$dinger--Foldy equation was given in our publications. Our motivation was as follows. The start of this model has been given in [18]. In the papers [19, 20] the two-component version of equation
$$
i\partial_{t}f(x)=\sqrt{m^{2}-\Delta}f(x)
$$
was considered and called as the Sch\"odinger equation. The one-component version of this equation was suggested in [38] and was called in the literature as the spinless Salpeter equation. In the papers considering the L$\mathrm{\acute{e}}$vy flights, see, e. g., [35], the spinless Salpeter equation is called as the L$\mathrm{\acute{e}}$vy-Schr$\mathrm{\ddot{o}}$dinger equation. Thus, the one-component case is identified well. But what about many-component case?

Taking into account the L. Foldy's contribution [18--20] in the construction of 2- and 4-component relativistic canonical quantum mechanics (RCQM) and his proof of the principle of correspondence between the relativistic and non-relativistic quantum mechanics, we propose \textit{to call the $N$-component equations of this type as the Schr$\mathrm{\ddot{o}}$dinger-Foldy equations}.

J.D. Bjorken and S.D. Drell in their well-known book [59] added some words to the original derivation [1], which make it possible to extend the derivation to $N$-dimensional case and, further in [60], to the $N$-dimensional space-time.

\textbf{2. Van der Waerden--Sakurai derivation.} In this version of the Dirac equation derivation the spin of the electron is incorporated into the nonrelativistic theory, see, e. g., [61]. Above considered Dirac's derivation was supported by the application of the relation $(\overrightarrow{\sigma} \cdot \widehat{\overrightarrow{p}})(\overrightarrow{\sigma} \cdot \widehat{\overrightarrow{p}})=\widehat{\overrightarrow{p}}^{2}$ and the factorization procedure in the form $(\widehat{p}^{0}+\overrightarrow{\sigma} \cdot \widehat{\overrightarrow{p}})(\widehat{p}^{0}-\overrightarrow{\sigma} \cdot \widehat{\overrightarrow{p}})=m^{2}$ of Van der Waerden is used. (Here $\widehat{\overrightarrow{p}}\equiv(\widehat{p}^{j})=-i\nabla$ is the quantum-mechanical momentum operator, $\widehat{p}^{0}\equiv i\partial_{0}$ is the operator of energy and $\overrightarrow{\sigma}$ are the Pauli matrices in the standard representation
 \begin{equation}
\label{eq1} \sigma^{1}=\left| {{\begin{array}{*{20}c}
 0 \hfill &  1 \hfill\\
 1 \hfill & 0  \hfill\\
 \end{array} }} \right|, \quad \sigma^{2}=\left| {{\begin{array}{*{20}c}
 0 \hfill &  -i \hfill\\
 i \hfill &   0  \hfill\\
 \end{array} }} \right|, \quad \sigma^{3}=\left| {{\begin{array}{*{20}c}
 1 \hfill &  0 \hfill\\
 0 \hfill & -1  \hfill\\
\end{array} }} \right|, \quad \sigma^{1}\sigma^{2}=i\sigma^{3}.
\end{equation}

Of course, the corresponded wave function in the equation
$$(\widehat{p}^{0}+\overrightarrow{\sigma} \cdot \widehat{\overrightarrow{p}})(\widehat{p}^{0}-\overrightarrow{\sigma} \cdot \widehat{\overrightarrow{p}})\phi=m^{2}\phi
$$
is a two-component. The procedure of transition from 2-component to 4-component wave function was fulfilled in [61] on the basis of analogy with electrodynamics, where the transition from 4-vector potential $A_\mu$ to the electromagnetic field tensor $F_{\mu\nu}(\overrightarrow{E},\overrightarrow{H})$ increases the number of components. Finally, the Dirac equation was derived. The details can be found in the monograph [61].

\textbf{3. Briefly on "factorization" of the Klein--Gordon operator.} In the well-known \textit{N.N. Bogoliubov and D.V. Shirkov book} [62] one can find a review of the Dirac theory, which is useful even today, and two different ways of the Dirac equation derivation. First, it is the presentation of the Klein--Gordon equation in the form of a first-order differential system of equations, i.e. so-called "factorization" of the Klein--Gordon operator in contemporary form. Following Dirac's publications [1, 17] and [61] authors demonstrated that
\begin{equation}
\Box -m^2 = (i\gamma^{\nu}\partial_\nu+m)(i\gamma^{\mu}\partial_\mu-m),
\label{eq2}
\end{equation}
\noindent where the matrix operators $\gamma^{\nu}$
\begin{equation}
\label{eq3} \gamma^{0}=\left| {{\begin{array}{*{20}c}
 \mathrm{I} \hfill &  0 \hfill\\
 0 \hfill & -\mathrm{I}  \hfill\\
 \end{array} }} \right|, \quad \gamma^{\ell}=\left| {{\begin{array}{*{20}c}
 0 \hfill &  \sigma^{\ell} \hfill\\
 -\sigma^{\ell} \hfill & 0  \hfill\\
 \end{array} }} \right|, \quad \ell=1,2,3,
\end{equation}
obey the conditions $\gamma^{\nu}\gamma^{\mu}+\gamma^{\mu}\gamma^{\nu}=2g^{\mu\nu}$ and $g^{\mu\nu}$ is the corresponding metric tensor. Assertion (2) is the main step of the given in [62] Dirac equation derivation.

\textbf{4. Principle of least action for the spinor field.} Application of the least action principle to the spinor field presents the next independent variant of the  Dirac equation derivation. Again one can refer to [62], where the Lagrange approach is considered and the Dirac equation is derived from the \textbf{variational Euler--Lagrange least action principle}.

The Lagrange function, for which the Euler--Lagrange equations coincide with the free Dirac equation
\begin{equation}
\label{eq4}
(i\gamma^{\mu}\partial_{\mu}-m)\psi(x)=0,
\end{equation}
is given by
\begin{equation}
\mathcal{L}_{\Psi}(x) = \frac{i}{2}\left(\overline{\psi}(x)\gamma^{\mu}\frac{\partial\psi}{\partial x^{\mu}} - \frac{\partial\overline{\psi}}{\partial x^{\mu}}\gamma^{\mu}\psi(x) \right)-m\overline{\psi}(x)\psi(x),
\label{eq5}
\end{equation}
\noindent where $\overline{\psi}(x)$ is the independent variational variable (in the set of solutions of the Dirac equation (4) $\overline{\psi}(x)=\psi^{\dagger}(x)\gamma^{0}$), symbol $"\dag"$  denotes the Hermitian conjugation.

The Dirac equation in external electromagnetic field of potentials $A^{\mu}(x)$ follows from the principle of least action, in which the Lagrange function have the form $\mathcal{L}=\mathcal{L}_{\Psi}+\mathcal{L}_{\mathrm{Int}}$, where the interaction Lagrangian is given by
\begin{equation}
\mathcal{L}_{\mathrm{Int}}(x) = e\overline{\psi}(x)\gamma^{\mu}\psi(x)A_{\mu}(x).
\label{eq6}
\end{equation}

\textbf{5. Ryder's derivation.} In the \textit{L. Ryder book} [63] (second edition) the Dirac equation is
derived \textbf{from the manifestly covariant transformational properties of the 4-component spinor under the Lorentz group}. This long derivation started from the rotation group and SU(2). After that the Lorentz group and SL(2,C) were considered. The transition from 2-spinors to the 4-spinors was caused by taking into account of parity property. Indeed, the $(j,0))$ and $(0,j)$ representations become interchanged $(j,0))\leftrightarrow(0,j)$ under parity. Further, on the basis of the Lorentz boost transformation and the supposition that the original spinor refers to the particle at rest in the form of the corresponding mathematica assertions ((2.85), (2.86) from [63]), author comes to the equation (2.93) in [63], which is the Weyl form of the Dirac equation. The details are presented in the book [63].

\textbf{6. Start from the initial geometric properties of the space-time and electron.} The derivation of the Dirac equation from the initial
geometric properties of the space-time and electron together with wide-range discussion of the geometric principles of the electron
theory is the main content of the \textit{J. Keller book} [64]. The ideas of \textit{V. Fock
and D. Iwanenko} [65, 66] on the geometrical sense of Dirac $\gamma$-matrices are the basis of the approach.

\textbf{7. Origin in Bargman--Wigner classification.} One should point out the derivation of the Dirac equation
based on the Bargman--Wigner classification of the irreducible unitary representations of the Poincar$\mathrm{\acute{e}}$ group, see, e.
g., [67, 68]. It is an illustrative demonstration of the possibilities of the group-theoretical approach to the elementary particle
physics. Indeed, within the group-theoretic approach to the elementary particles physics, the Dirac equation follows from the Bargman--Wigner classification [67, 68] of the elementary particles of arbitrary mass and spin on the basis of irreducible unitary representations of the Poincar$\mathrm{\acute{e}}$ group. The Dirac Hamiltonian corresponds to the irreducible representation of this group, which is characterized by the eigen values $m>0$ and $s=1/2$ of the corresponded Casimir operators.

\textbf{8. The partial case of the Bargman--Wigner equation.} Dirac equation is the partial case of the Bargman--Wigner equation [68] for the arbitrary spin fields. Indeed, after the substitution s=1/2 case into 2s$\cdot$4-component Bargman--Wigner equation [68] one comes to 4-component Dirac equation. Nevertheless, such derivation is week. Note that Bargman--Wigner equation has been introduced as the 2s$\cdot$4-component generalization of the Dirac equation.

Similarly (as spin 1/2 partial cases), the Dirac equation can be derived from the Bhabha, Pauli--Fierz, Rarita--Schwinger equations, or from other equations for arbitrary spin. Such weak derivations are not considered here.

\textbf{9. Inverse Foldy--Wouthuysen transformation.} In \textit{L. Foldy and S. Wouthuysen} paper [18], and in \textit{L. Foldy} articles [19, 20], one can easy find the inverse problem, in which the Dirac equation can be derived from the canonical FW equation.

Note that the FW transformation was introduced in the Dirac theory in order to transform the Dirac equation into the special canonical form, in which the quantum-mechanical interpretation is more evident in comparison with standard manifestly covariant form. Indeed, in the FW representation (contrary to the Dirac representation) the operators of position and velocity are defined well and the spin commutes with the Hamiltonian itself (not only together with orbital angular momentum). The orbital angular momentum commutes here with the Hamiltonian as well. The FW representation is considered in details in the monograph [3] (Chapters 4, 6).

However, in the FW representation the transition to the proper quantum mechanics is not finished. The operators of energy and spin in the FW representation still can not be interpreted in the framework of quantum mechanics. The operator of energy leads to the solutions with positive  and negative energies and spin projection for the particle and antiparticle is similar. The well-defined energy and spin operators are obtained after the extended FW transformation, which translates the Dirac model into the well-defined quantum-mechanical model without negative energies and with different projections of spin for the particle and antiparticle. Such extended transformation has been suggested in our publications [3, 15, 16, 56--58] and the corresponding representation of the Dirac theory has been called as the RCQM. This model  and its relation with the Dirac theory is considered below in Item 33.

Not a matter of fact the FW representation is well-known, see, e.g., [21--37] and the references therein. Therefore, the derivation of the Dirac equation from the FW equation must be considered in the list of Dirac equation derivations. Nevertheless, it is only the transition from one representation of the spinor field to another, from the close to quantum mechanics nonlocal canonical representation to the local field manifestly covariant picture.

\textbf{10. Feynman's derivation.} The intriguing derivation was suggested by Feynman (for the SO(1,1) case with one spatial dimension and one time dimension) and was given as a problem in his book with Hibbs [69]. This approach was continued in [70], where the path integral constitutes a generalization of Feynman's checkerboard model to $3+1$ space-time dimensions.

\begin{center}
\textbf{Unexpected derivations}
\end{center}

\textbf{11. Sallhofer's derivation from the Maxwell electrodynamics in medium.} The first one who should be mentioned is the austrian scientist H. Sallhofer. The following assertion has been proved [71, 72]. \textit{Matrix multiplication of the Maxwell equations in medium (without currents and charges) on the Pauli vector $\vec{\sigma}=(\sigma^{1},\sigma^{2},\sigma^{3})$ gives the massless stationary Dirac equation}. Not stopping at this, but taking this result as a basis, he put into consideration for the electric and magnetic permeabilities of the inner-atomic medium the representation as follows (here $\hbar\neq c\neq 1$)
\begin{equation}
\varepsilon \left(\overrightarrow{x}\right) =1-\frac{\Phi \left(
\overrightarrow{x}\right) -m c^{2}}{\hbar \widetilde{\omega}},\quad \mu \left(
\overrightarrow{x}\right) =1-\frac{\Phi \left( \overrightarrow{x}\right) +
m c^{2}}{\hbar \widetilde{\omega}},
\label{eq7}
\end{equation}
\noindent where $\Phi \left(\overrightarrow{x}\right)=-Ze^2/\left|\overrightarrow{x}\right|$ is the external Coulomb field, $\hbar$ is the Planck constant and $c$ is the velocity of light. Further, the Sommerfeld--Dirac formula for the hydrogen spectrum fine structure has been derived [73, 74] from the Maxwell equations in medium (7) (for the first time after Dirac from a fundamentally different equation and even without appealing to quantum mechanics). Thus, papers [71--74] contain considerably more information than the simple derivation of the Dirac equation from the Maxwell equations.

\textbf{12. In addition to Ryder's derivation.} Some remarks about Ryder's derivation of the Dirac equation were considered in [75], where the formalism of [63] was repeated together with taking into account the physical meaning of the negative energies and the relative intrinsic parity of the elementary particles. The authors believe that the derivation from [63] is improved.

\textbf{13. Derivation from the Langevin equation for a two-valued process.} The goal of contribution [76] was to show that the relativistic equation for spin-1/2 particles can be obtained by an enlargement of the theory of stochastic processes one step beyond the theory of complex measures. The quaternion measurable processes were introduced and the Dirac equation was derived from the Langevin equation associated with a two-valued process. A direct derivation of the Dirac equation via quaternion measures has been suggested.

Authors introduced the quaternion measure similarly to the complex measure, then they were able to define Markov processes $X(t)$ and derived the Chapman--Kolmogorov relation. Within such a framework of quaternion measurable processes, the Langevin equation has been considered
\begin{equation}
dx^{j}=v^{j}\exp[i\pi z(t)]dt,
\label{eq8}
\end{equation}
where $v^{j}$ is a function of $x$ and $z(t)$ is a two-valued Markov process on 0,1 with transition rates $\lambda_{+}(0\rightarrow 1)$ and $\lambda_{-}(1\rightarrow 0)$ per unit time. Thus, the process $z(t)$ represents the transition from one to the other of the two helicity states.

After some additional assumptions and suggestions in [76] authors adapted the Fokker--Plank equation (used before in [76] in derivation of the Pauli equation) to yield
\begin{equation}
\frac{\partial \pi_{+}(x,t)}{\partial t}=-\sigma \nabla \pi_{+}-(\lambda_{+}\pi_{+}-\lambda_{-}\pi_{-}),
\label{eq9}
\end{equation}
where $\pi_{+}(x,t)dx\pi_{-}(x,t)$ represents quaternion measure that $x(t)$ lies in $(x, \, x+dx)$ and $z(t)=0 \, (z(t)=1)$. In a similar way the equation
\begin{equation}
\frac{\partial \pi_{-}(x,t)}{\partial t}=+\sigma \nabla \pi_{-}-(\lambda_{-}\pi_{-}-\lambda_{+}\pi_{+})
\label{eq10}
\end{equation}
was obtained.
After using the postmultiplication of $\pi_{\pm}$ by an arbitrary 2-spinors, misnotation of $\pi_{\pm}$, choosing $\lambda_{\pm}$ and satisfying
\begin{equation}
\psi=\left|
\begin{array}{cccc}
 \pi_{+} \\
 \pi_{-} \\
 \end{array} \right| e^-{imt}
\label{eq11}
\end{equation}
authors successfully obtained the Dirac equation in the Weyl representation.

\textbf{14. Derivation from the conservation law of spin 1/2 current.} The author of [77] was able to derive the Dirac equation from the conservation law of spin 1/2 current. The requirement that this current is conserved leads to a unique determination of the Lorentz invariant equation satisfied by the relativistic spin 1/2 field. If so, then what is distinguished the conservation law of spin-current in comparison with other constants of motion? Therefore, it is logical to expect the successful derivations of the Dirac equation from other conservation laws of the spinor field.

Let us briefly comment that the complete list of fundamental conservation laws for the spinor field is the Noether consequence of the Dirac equation and its symmetries. The 10 main conserved quantities are the consequences of the Poincar$\mathrm{\acute{e}}$ symmetry. Moreover, different additional and so-called hidden symmetries are known, see, e. g., [9--14]. The additional and hidden conservation laws are known for the spinor field as well [12--14], where not only the Fermi but also the Bose constants of motion are found. Note especially the list of additional conservation laws for the spinor field [78], which was found in the FW representation on the basis of the Noether theorem.

Moreover, as soon as the conserved currents follow directly from some equation of motion, therefore, the validity of the inverse problem is really expected. And not only spin 1/2 conserved current can be useful.

\textbf{15. Derivation from the master equation of Poisson process.} The Dirac equation was derived [79] from the master equation of Poisson process by analytic continuation. The extension to the case, where a particle moves in an external field, was given. It was shown that the generalized master equation is closely related to the three-dimensional Dirac equation in an external field.

In this brief letter authors  extended the path-integral formulation of the Poisson process and take into account the effect of the external field in the one-dimensional framework, and furthermore, in the three-dimensional version. Thus, firstly, the derivation of the one-dimensional Dirac equation $i\partial_{t}\phi=m\sigma_{1}\phi -i\sigma_{3}\partial_{x}\phi+V(t)\phi$ from the Poisson process was given. After that the generalization for three spatial dimensions has been  considered.

\textbf{16. Derivation from the relativistic Newton's second law.} In [80], a method of deriving the Dirac equation from the relativistic Newton's second law was suggested. However, for these purposes the author put into consideration the list of his own definitions, concepts and, even, special models. Indeed, this derivation is possible in a new formalism, which relates the special form of relativistic mechanics to the quantum mechanics. The author suggested a concept of a velocity field. At first, the relativistic Newton's second law was rewritten as a field equation in terms of the velocity field, which directly reveals a new relationship linked to the quantum mechanics. After that it was shown that the Dirac equation can be derived from the field equation in a rigorous and consistent manner. The impression is that the author seeks for coordination of his approach with the well-tested model of Dirac.

\textbf{17. A geometrical derivation of the Dirac equation.} A geometrical derivation of the Dirac equation, by considering a spin 1/2 particle traveling with the speed of light in a cubic space-time lattice, was made in [81]. The mass of the particle acts to flip the multi-component wave function at the lattice sites. Starting with a difference equation for the case of one spatial and one time dimensions, the authors generalize the approach to higher dimensions. Interactions with external electromagnetic and gravitational fields are also considered. Nevertheless, the idea of such derivation is based on the Dirac's observation that the instantaneous velocity operators of the spin 1/2 particle (hereafter called by the generic name "the electron") have eigenvalues $\pm c$. Note that today this fact is considered as a difficulty in the Dirac equation quantum-mechanical interpretation and was demonstrated, explained and overcome in the paper [18] (see the Chapter 4 in the monograph [3] for some additional details).

It will be useful to consider the link of this approach with the geometric consideration of V. Fock, D. Iwanenko, J. Keller [64--66] briefly presented here in Subsection 6. Unfortunately, in [81] such link is absent. The appealing to such assumption as a $\pm c$ velocity of the massive fermion is a shortcoming of the formalism of [81].

\textbf{18. Start from the geodesic equation.} Using the mathematical tool of Hamilton's bi-quaternions, the authors of [82] propose a derivation of the Dirac equation from the geodesic equation. Such derivation is given in the program of application of the theory of scale relativity to the purposes of microphysics at recovering quantum mechanics as a new non-classical mechanics on a non-derivable space-time.

\textbf{19. Derivation from a generally covariant field equation for gravitation and electromagnetism.} M. Evans was successful to express his equation of general relativity (generally covariant field equation for gravitation and electromagnetism [83]) in spinor form, thus producing the Dirac equation in general relativity [84]. The Dirac equation in special relativity is recovered in the limit of Euclidean or flat space-time. Thus, the Dirac equation was derived from a generally covariant field equation for gravitation and electromagnetism.

\textbf{20. Derivation from the unique conditions for wave function.} Author of [85] first determines that each eigenfunction of a bound particle is a specific superposition of plane wave states that fulfills the averaged energy relation. After that the Schrodinger and Dirac equations were derived as the unique conditions the wave function must satisfy at each point in order to fulfill the corresponding energy equation. The Dirac equation involving electromagnetic potentials has been derived.

\textbf{21. Derivation by factorizing the algebraic relation satisfied by the classical Hamiltonian.} Author of this approach [86] started from classical-quantum correspondence, that associates a linear differential operator with a classical Hamiltonian, and that leads to regard energy and momentum as operators of this kind. The Dirac equation is derived by factorizing the algebraic relation satisfied by the classical Hamiltonian. The Whitham consideration [87] of the theory of classical waves was used, where for any linear wave equation one may define different "`wave modes"', each of which is characterized by a \textit{dispersion relation}, i.e., an explicit dependence of the frequency $\omega$ as a function of the spatial wave (co-)vector \textbf{k}, $\omega = W(\textbf{k}; X)$ (in the general case of heterogeneous propagation, when the dispersion depends indeed on the space-time position $X$). It turns out [87] that for a given wave mode the wave vector \textbf{k} propagates along the bicharacteristics of a certain linear partial differential equation of the first order. When the latter equation is put into characteristic form, one obtains a \textit{Hamiltonian system}, in which the Hamiltonian is none other than the dispersion relation \textit{W} defining the given wave mode of the wave equation considered. Therefore, with certain precautions, which are made necessary by the existence of several wave modes, one may \textit{recover the wave equation} from the dispersion relation alone. It is the basis of the derivation of the Dirac equation in [86] for the free particle, for the case of external electromagnetic field and for the case of static gravitational field. Note that the main point in such derivation is near standard factorization of the Klein--Gordon operator considered above in Item 3.

\textbf{22. Origin in conformal differential geometry.} In [88] the Dirac equation is derived by conformal differential geometry. The Hamilton--Jacobi equation for the particle is found to be linearized, exactly and in closed form, by an ansatz solution that can be straightforwardly interpreted as the quantum wave function of the 4-spinor solution of Dirac's equation. All quantum features arise from the subtle interplay between the conformal curvature acting on the particle as a potential and the particle motion which affects the geometric prepotential associated to the conformal curvature itself. Finally, the Dirac equation is found in the form $\widehat{D}_{+}\widehat{D}_{-}\psi=\widehat{D}_{-}\widehat{D}_{+}\psi=0, \, \, \widehat{D}_{\mp} \equiv \gamma^{\mu}\left(p_{\mu}-eA_{\mu}\right)\mp m$.

\textbf{23. Derivation from principles of information processing.} In the article [89] the derivation of the Dirac equation from principles of information processing has been presented. It has been shown, without using the relativity principle, how the Dirac equation in three space-dimensions emerges from the large-scale dynamics of the minimal nontrivial quantum cellular automaton satisfying unitarity, locality, homogeneity, and discrete isotropy. The Dirac equation is recovered for small wave-vector and inertial mass, whereas Lorentz covariance is distorted in the ultra-relativistic limit. The automaton can thus be regarded as a theory unifying scales from Planck to Fermi.

\textbf{24. A method to derive the anticommutation properties of the Dirac matrices without relying on the squaring of the Dirac Hamiltonian.} In original approach [90] the anticommutation properties of the Dirac matrices can be derived without squaring the Dirac Hamiltonian, that is, without any explicit reference to the Klein--Gordon equation. The only requirement is as follows. The Dirac equation is considered to admit two linearly independent plane wave solutions with positive energy for all momenta. The necessity of negative energies as well as the trace and determinant properties of the Dirac matrices are also a direct consequence of this simple and minimal requirement. Nevertheless, necessity of negative energies is not the positive feature of the formalism.

\textbf{25. Equation of motion for the spherical spin model.} In the paper [91] the Dirac equation was derived as the equation of motion for the spherical spin model. An attempt was made to show that fundamental particles are manifestations of the geometry of space-time. This was donee by demonstrating the existence of a purely geometrical model, which was called spherical rotation, that satisfies Dirac's equation. The model was developed and illustrated both mathematically and mechanically. It indicates that the mass of a particle is entirely due to the spinning of the space-time continuum. Using the model, authors showed the distinction between spin-up and spin-down states and also between particle and
antiparticle states. It satisfies Einstein's criteria for a model that has both wave and particle properties, and it does so without introducing a singularity into the continuum.

\textbf{26. Dirac equation from the strands and tangles.} Recently in the publication [92] the wave functions, the Dirac equation, the least action principle and the Lagrangian for a free fermion were deduced from strands. The author's conjecture is as follows. In flat space, fluctuating strands forming rational tangles, i.e., unknotted tangles, yield a model for elementary particles. The conjecture derives from Dirac's proposal to describe fermions as tethered objects and models elementary particles as rational tangles. The tangle model appears to explain the Dirac equation, the principle of least action, the observed particle spectrum of fermions and bosons, and the three observed gauge interactions with their Lie groups and all their other properties. In a natural way, the specific tangles for each elementary particle define spin, quantum numbers and all other properties. In some sense, these are continuations of Battey-Pratt and Racey ideas [91].

\textbf{27. A diffusion model for the Dirac equation.}

Using a physical model that featured a special type of diffusion process, the author of [93] was able to derive the Sch\"odinger equation from mass and momentum conservation equations associated with this diffusion process. After that in the  paper [94] an attempt is made to extend the treatment of [93] to include the relativistic wave equations, in particular, the Dirac equation. Again, the treatment begins by considering the basic physical model. However, for the Dirac equation, a new feature must be added to this model in the form of the intrinsic magnetic and electric dipoles that are to be associated with each particle.

Finally, the author was able to derive the Dirac equation from the constructed model. The author claims that deriving the Dirac equation with the foregoing treatment is longer, more involved, and more inelegant, mathematically speaking, than the type of derivation first used by Dirac and now generally used in quantum mechanics. It is claimed here, however, that the present treatment has a significant advantage (at least for some) in that it is physically understandable at every step and does not employ what (to the practically inclined) might seem to be vague mathematical artifices. However, the article [94] was cited only five times.

\textbf{28. A variational principle for the Feynman's derivation.}

In addition to Feynman's path integral derivation from Item 10 in the paper [95] the variational principle and corresponding Euler--Lagrange equation for the ordinary 4-component Dirac equation was suggested. As a preliminary result the one-dimensional Dirac equation in the Weyl representation
\begin{equation}
i\partial_{t}\phi+i\sigma_{3}\partial_{x}\phi=m\sigma_{1}\phi
\label{eq12}
\end{equation}
was derived. Here $\sigma_{1}, \sigma_{3}$ are the standard $2\times 2$ Pauli matrices and the two components of $\phi=\left|
\begin{array}{cccc}
 \phi^{+} \\
 \phi^{-} \\
 \end{array} \right|$ describe, as $m\rightarrow 0$, respectively a right-handed (spin parallel to the momentum direction) and the left-handed (spin antiparallel to the momentum direction) state of the spin-$\frac{1}{2}$ fermion.

On this basis the generalisation of the least action principle for the Minkowski space-time and the derivation of the ordinary 4-component Dirac equation was considered.

\textbf{29. The Dirac equation from scratch.} The derivation of the free-space Dirac equation from scratch has been discussed in the book [96],
especially in the Chapter 5, and in the HAL arxive [97]. Author started from the assumption that an electron at rest spins with an angular frequency $\omega_{0}$ around a fixed spin-axis is defined by a unit vector $\overrightarrow{n}\in \mathbb{R}^{3}$. The essential step is based at the Rodrigues formula application:
\begin{equation}
R(\overrightarrow{n},\phi)=\cos \frac{\phi}{2}\mathbb{I}-i\overrightarrow{n}\cdot \overrightarrow{\sigma}\sin \frac{\phi}{2},
\label{eq13}
\end{equation}
where $R(\overrightarrow{n},\phi)$ is a rotation over an angle $\phi$ around an axis with unit vector $\overrightarrow{n}$. Further using some assumptions like $\frac{\hbar \omega_{0}}{2}=m_{0}c^{2}$, identities as $[\overrightarrow{n}\cdot \overrightarrow{\sigma}]^{2}=\mathbb{I}$ and analogy between some geometric and some Dirac formulas author is able to derive the Dirac equation.

\textbf{30. Another attempt from scratch.} Near centenary after the Heisenberg's, Schr$\mathrm{\ddot{o}}$dinger's, Klein--Gordon's and Dirac's equations derivations many authors marked that these approaches were based on heuristics in getting from classical mechanic Hamiltonians to quantum mechanic equations of motions. The assertion that these derivations did not started from basic physical principles besides classical mechanics was considered by many authors.

In the brief note [98] an attempt to suggest the original point of view was given. It was based on the analysis of the Dirac vector space in a mathematical rigid manner, the physical conditions of Lorentz invariance and positive definiteness of the Dirac's term $\overline{\Psi}\Psi$. By doing this author had come to the Dirac equation in momentum space with the gamma matrices in the chiral or Weyl representation. The weaknesses are as follows. According to the author's own remark after this derivation there are no massless spin 1/2 fermions in the Dirac formalism because a nontrivial mass is essential for the derivation. Moreover, the Dirac's term $\overline{\Psi}\Psi$, as well as the Dirac conjugation in general,  was unknown before the existing of Dirac equation and his spinor field. Therefore, such derivation in any case is not from scratch.

Nota bene! Many "modern methods" of the Dirac equation derivation are impossible in principle without the preliminary knowledge of his equation, corresponding Dirac's suggestions and spinor field formalism. Thus, the last remark is related to many investigators of the Dirac equation derivation, not only to the author of this brief communication in [98]. We ask authors to find such methodological kinds of their errors themselves.

\textbf{31. Scale-relativistic derivations of Pauli and Dirac equations.} The author of [99] considered so called scale relativity. In this model the quantum mechanics is recovered by transcribing the classical equations of motion to fractal spaces and demanding, as dictated by the principle of scale relativity, that the form of these equations be preserved. In the framework of this approach, however, the form of the classical energy
equations both in the relativistic and nonrelativistic cases are not preserved. Aiming to get full covariance, i.e., to restore to these equations their classical forms, the author demonstrated that the scale-relativistic form of the Schr$\mathrm{\ddot{o}}$dinger equation yields the Pauli equation, whilst the Pissondes's scale-relativistic form of the Klein--Gordon equation gives the Dirac equation.

The Pissondes's form of the Klein--Gordon equation is given by
\begin{equation}
\mathcal{V}^{\mu}\mathcal{V}_{\mu}+2iK\partial^{\mu}\mathcal{V}_{\mu}-1=0,
\label{eq14}
\end{equation}
where
\begin{equation}
\mathcal{V}^{\mu}=\frac{i}{m}(\partial^{\mu}\psi) \psi^{-1}, \quad K=\frac{1}{2m},
\label{eq15}
\end{equation}
and other notations of the model of scale relativity are used. Further, the procedure of factorising the Klein--Gordon equation (14) was applied.

\textbf{32. Recent derivation of Dirac equation from the stochastic optimal control principles of quantum mechanics.} In the paper [100], author presents a stochastic approach to relativistic quantum mechanics. The three fundamental principles of this model have been formulated. The Dirac equation was derived on the basis of these principles. This approach enables readers to bring more insight into the nature of Dirac's spinors. Furthermore, a physical interpretation of the stochastic optimal control theory of quantum mechanics has been suggested.

Three main principles of the model construction are formulated as follows. (i) Particles move as Brownian particles in four-dimensional spacetime, influenced by an external random spacetime force. (ii) The stochastic movement of the particle turns its classical action integral into a stochastic variable. (iii) Nature tries to minimize the expected value for the action, in which the particle's velocity is consider to be a control parameter of the optimization.

Note that these principles were already put into consideration by the authors of the foundations of stochastic quantum mechanics and optimal control of the corresponding model.

On this basis author of the paper [100] was able to derive the Dirac equation by linearizing the stochastic Hamilton--Jacobi--Bellman equations, which is a partial differential equation with boundary conditions. More directly, the idea of the linearization of the Lagrangian in this equation was the main idea in this paper, which was realized in the derivation of the Dirac equation.

\begin{center}
\textbf{Our approaches to the problem of the Dirac equation derivation}
\end{center}

We have contributed in the methods of Dirac equation derivation thrice: (i) the Sallhofer's derivation from the Maxwell electrodynamics in medium was generalized to the case of non-stationary equation and the complete set of corresponding transformations was found, (ii) the massless Dirac equation was derived from the maximally symmetrical form of the Maxwell equations, (iii) the Dirac equation with nonzero mass was derived from the relativistic canonical quantum mechanics (RCQM) of spin 1/2 particle-antiparticle doublet. It determines our interest to the problem, in which one of our goals is to find our place among the other authors.

\textbf{33. Generalization of Sallhofer's derivation.} As it is already presented in Item 11 above, in the papers [71--74] the link between the Dirac equation in external electromagnetic field and the Maxwell equations in medium has been given. In the paper [101] Sallhofer's idea was generalized to the case of non-stationary Dirac equation and was presented in the framework of another formalism. Moreover, in [101] the complete set of linking transformation was found. The main assertion of the publication [101] is as follows.

Maxwell's equations of source-free electrodynamics
\begin{equation}
\frac{\varepsilon}{c}\partial_{0}\overrightarrow{E}-\mathrm{curl}\overrightarrow{H}=0, \, \frac{\mu}{c}\partial_{0}\overrightarrow{H}+\mathrm{curl}\overrightarrow{E}=0, \quad \mathrm{div}\overrightarrow{E}=0, \, \mathrm{div}\overrightarrow{H}=0,
\label{eq16}
\end{equation}
\noindent ($c$ is the light velocity, $\varepsilon $ and $\mu $ are the electric and magnetic permeabilities of the medium) are linked to the Dirac-like equation
\begin{equation}
\left[\overrightarrow{\alpha}\cdot\nabla-\left|
{{\begin{array}{*{20}c}
 \varepsilon \mathrm{I_{2}} \hfill & 0 \\
 0 \hfill & \mu \mathrm{I_{2}} \\
\end{array} }} \right|\frac{1}{c}\frac{\partial}{\partial t}\right]\psi^{\mathrm{el}}=0,
\label{eq17}
\end{equation}
\noindent where $\overrightarrow{\alpha}=\left|
{{\begin{array}{*{20}c}
 0 \hfill & \overrightarrow{\sigma} \\
 \overrightarrow{\sigma} \hfill & 0  \\
\end{array} }} \right|$, $\overrightarrow{\sigma}=\left\{\sigma^{1},\sigma^{2},\sigma^{3}\right\}$ are the Pauli matrices (1), $\psi^{\mathrm{el}}$ is one of eight columns known from the paper [101]
\begin{equation}
\psi^{\mathrm{el}} = \left|
\begin{array}{cccc}
 iE^{3} \\
 i(E^{1}+iE^{2}) \\
 H^{3} \\
 H^{1}+iH^{2} \\
 \end{array} \right|,
\label{eq18}
\end{equation}
\noindent  $\mathrm{I_{2}}$ is $2 \times 2$ unit matrix (In (16), (18) and below ($\overrightarrow{E},\overrightarrow{H}$) are the electromagnetic field strengths).

By applying the matrix-differential operator of (17) to the column (18) we obtain a system of four equations, in which the imaginary and real parts can easily be separated. Then, by natural requiring that the imaginary and real parts be equal to zero independently, we come to Maxwell's equations (16).

It is evident that column $\psi^{\mathrm{el}}$ (16) contains only six real functions ($\overrightarrow{E},\overrightarrow{H}$), while four-component Dirac spinor $\psi^{\mathrm{D}}$ contains eight similar real functions, which in general are not related to ($\overrightarrow{E},\overrightarrow{H}$). Therefore, the components of the column $\psi^{\mathrm{el}}$ (18) are the subset of the components of Dirac spinor $\psi^{\mathrm{D}}$. Hence, there is no one to one correspondence between components of $\psi^{\mathrm{el}}$ (18) and $\psi^{\mathrm{D}}$.

Therefore, the inverse assertion that the massless Dirac equation follows from the Maxwell equations (16) is not so evident. In order to fulfill such inverse derivation it is necessary to put $\varepsilon=\mu=1$ and to complete the 6 real components of (18) to the 8 real components of the standard Dirac spinor.

The complete set of columns, which can be chosen as $\psi^{\mathrm{el}}$ in (18), is eight. In the notations
\begin{equation}
E^{\pm} = E^{1}\pm iE^{2}, \quad H^{\pm} = H^{1}\pm iH^{2},
\label{eq19}
\end{equation}
\noindent they are given by:
\begin{equation}
\psi^{1} = \left|
\begin{array}{cccc}
 iE^{3} \\
 iE^{+} \\
 H^{3} \\
 H^{+} \\
 \end{array} \right|, \,
\psi^{2} = \left|
\begin{array}{cccc}
 -E^{3} \\
 -E^{+} \\
 iH^{3} \\
 iH^{+} \\
 \end{array} \right|, \,
\psi^{3} = \left|
\begin{array}{cccc}
 H^{3} \\
 H^{+} \\
 iE^{3} \\
 iE^{+} \\
 \end{array} \right|, \,
\psi^{4} = \left|
\begin{array}{cccc}
 iH^{3} \\
 iH^{+} \\
 -E^{3} \\
 -E^{+} \\
 \end{array} \right|,
\label{eq20}
\end{equation}
$$\psi^{5} = \left|
\begin{array}{cccc}
 -iH^{-} \\
 iH^{3} \\
 E^{-} \\
 -E^{3} \\
 \end{array} \right|, \,
\psi^{6} = \left|
\begin{array}{cccc}
 H^{-} \\
 -H^{3} \\
 iE^{-} \\
 -iE^{3} \\
 \end{array} \right|, \,
\psi^{7} = \left|
\begin{array}{cccc}
 E^{-} \\
 -E^{3} \\
 -iH^{-} \\
 iH^{3} \\
 \end{array} \right|, \,
\psi^{8} = \left|
\begin{array}{cccc}
 iE^{-} \\
 -iE^{3} \\
 H^{-} \\
 -H^{3} \\
 \end{array} \right|.$$
\noindent (The set (20) includes column (18) as $\psi^{1}$).

The validity and suitability of the set (20) can \textit{be proven} by direct substitution of every column of this set for $\psi^{\mathrm{el}}$ in the equation (17). In order to prove that it is really a complete set let us recall the eight Pauli--G$\ddot{\mathrm{u}}$rsey--Ibragimov operators [102--104]
 $$\{\gamma^{2}\hat{C}, \, i\gamma^{2}\hat{C}, \,
\gamma^{2}\gamma^{4}\hat{C}, \, i\gamma^{2}\gamma^{4}\hat{C}, \,
\gamma^{4},  \, i\gamma^{4}, \, i, \, \mathrm{I} \},
$$
 where ($\hat{C}$ is the $4\times4$ matrix operator of complex conjugation, $\hat{C} \psi = \psi^{*}$, the operator of involution in the Hilbert space $\mathrm{H}^{3,4}$). These generators form a complete set of pure matrix operators, which leave the massless Dirac equation being invariant. The verification that eight columns from the set (20) are obtained by action of every Pauli--G$\ddot{\mathrm{u}}$rsey--Ibragimov operator on the function $\psi^{\mathrm{el}}$ (18) finishes the proof. As soon as we have in $\{\gamma^{2}\hat{C}, \, i\gamma^{2}\hat{C}, \,
\gamma^{2}\gamma^{4}\hat{C}, \, i\gamma^{2}\gamma^{4}\hat{C}, \,
\gamma^{4},  \, i\gamma^{4}, \, i, \, \mathrm{I} \}$ the complete set of pure matrix operators, therefore, the (20) is the complete set too.

Note that the four columns $\psi^{3-6}$ can be chosen in (20) only together with simultaneous interchange $\varepsilon \leftrightarrow \mu$  in (17).

Note, further, that in [105--107] different useful matrix representations of the Lie algebras of the Lorentz and Poincar$\mathrm{\acute{e}}$ groups have been found on the basis of Pauli--G$\ddot{\mathrm{u}}$rsey--Ibragimov operators  [102--104]. At first, it is the additional D($0,\frac{1}{2})\oplus(\frac{1}{2},0$) representation of the Lie algebra of \textit{universal covering} $\mathcal{L}$ = SL(2,C) of the proper ortochronous Lorentz group $\mbox{L}_ + ^\uparrow $ = SO(1,3)=$\left\{\Lambda=\left(\Lambda^{\mu}_{\nu}\right)\right\}$. Further, the application of the simplest linear combinations  of both standard and additional D($0,\frac{1}{2})\oplus(\frac{1}{2},0$) generators gives the possibility to find Bose representations of the Lorentz group $\mathcal {L}$ and the Poincar$\mathrm{\acute{e}}$ group $\mathcal{P}\supset\mathcal{L}$ = SL(2,C) (here $\mathcal{P}$ is the universal covering of the proper ortochronous Poincar$\mathrm{\acute{e}}$ group $\mbox{P}_ + ^ \uparrow = \mbox{T(4)}\times )\mbox{L}_ + ^ \uparrow  \supset \mbox{L}_ + ^\uparrow$), with respect to which the massless Dirac equation is invariant. Thus, we have found [105--107] the Bose D(1,0)$\oplus$(0,0) and D$(\frac{1}{2},\frac{1}{2})$ representations of the Lie algebra of the Lorentz group $\mathcal{L}$ together with the tensor-scalar of the spin s=(1,0) and vector representations of the Lie algebra of the Poincar$\mathrm{\acute{e}}$ group $\mathcal{P}$, with respect to which the Dirac equation with m=0 is invariant.

Assuming harmonic time dependence in the form
\begin{equation}
\Psi =\psi e^{-i\widetilde{\omega} t} \rightarrow \frac{\partial}{\partial (ct)}\Psi=-i\frac{\widetilde{\omega}}{c}\Psi,
\label{eq21}
\end{equation}
\noindent from (17) the electromagnetic amplitude equation
\begin{equation}
\left[\overrightarrow{\alpha}\cdot\nabla+i\frac{\tilde{\omega}}{c}\left|
{{\begin{array}{*{20}c}
 \varepsilon \mathrm{I_{2}} \hfill & 0 \\
 0 \hfill & \mu \mathrm{I_{2}} \\
\end{array} }} \right|\right]\psi^{\mathrm{el}}=0
\label{eq22}
\end{equation}
follows. A comparison with the Dirac amplitude equation
\begin{equation}
\left[\overrightarrow{\alpha}\cdot\nabla+i\frac{\widetilde{\omega}}{c}\left|
{{\begin{array}{*{20}c}
 \left(1-\frac{\Phi \left(
\overrightarrow{x}\right) -mc^{2}}{\hbar \widetilde{\omega}}\right) \mathrm{I_{2}} \hfill & 0 \\
 0 \hfill & \left(1-\frac{\Phi \left( \overrightarrow{x}\right) +
mc^{2}}{\hbar \widetilde{\omega}}\right) \mathrm{I_{2}} \\
\end{array} }} \right|\right]\psi^{\mathrm{D}}=0
\label{eq23}
\end{equation}
in external Coulomb field $\Phi \left(\overrightarrow{x}\right)=-Ze^2/r$ immediately shows the complete formal agreement of electrodynamics with the Dirac theory. A consequence of such a comparison are the \textit{Sallhofer's} formulas (7), on the basis of which the description of hydrogen atom relativistic spectrum was presented in [73, 74] by means of (7) and (16), i.e., in terms of Maxwell's electrodynamics.

Similarly to the consideration above of the column $\psi^{\mathrm{el}}$ (18) and four-component Dirac spinor $\psi^{\mathrm{D}}$, here again there is absent the one to one correspondence between 6 real components of $\psi^{\mathrm{el}}$ (22) and 8 real components of $\psi^{\mathrm{D}}$ (23).

Not a matter of fact that above some formal agreement of electrodynamics with the Dirac theory is presented the fine structure of hydrogen atom spectrum on this basis has been found both with application of (16) [73, 74] and with application of 8-component form of Maxwell equations [106--108] . Such Maxwell system is considered in Item 32 below. Contrary to the formalism above the components of these slightly generalized Maxwell equations possess one to one correspondence with the Dirac theory.

It is better to consider the above given formalism from [101] together with our paper [109]. In [109] different relations between the free Maxwell equations (16), when $\varepsilon=\mu=1$, and the massless Dirac equations were investigated: The relationships between the Lagrangians, symmetries, solutions of free Maxwell equations and Lagrangians, symmetries, solutions of the massless Dirac equations were found. Furthermore, the relationships between the conservation laws for the electromagnetic and massless spinor field were found in [109] as well.

Note that recently the authors of [110] also marked the difference between the approaches in the papers [71, 72] and [101, 111, 112]. Moreover, author of [71, 72] named his model "The Maxwell-Dirac isomorphism". Note further that "isomorphism" in the papers [71--74], as it is evident from consideration above, can be considered rather as the physical, not as the mathematical assertion. In the papers [111--114] we named it (together with the case of free Maxwell equations in vacuum in [109]) as the "relationship" between the Dirac and the Maxwell equations. J. Keller in his review [115] named it the "mapping" of the Maxwell formalism into the Dirac formalism. In our derivations [71, 72, 101] both Johan Sallhofer and the author of this manuscript used such methods as the similarity and analogy in formulas and expressions, which are not well-defined mathematically. However, in the next investigations of the problem [3, 111--114] and [105--108] the natural and the exact mathematical proofs were given.

Note that recently the relationships between the Dirac and the Maxwell equations were reviewed in the paper [116]. Unfortunately, the author demonstrated a poor knowledge of the problem. The mistakes as "The Maxwell--Dirac isomorphism" started even in the title of the paper. How it can be revisited if it is not existing in mathematics and mathematical physics? The Maxwell and Dirac equations cannot be isomorphic. The reasons are the the different number of the components of the wave solutions and the different (zero and nonzero) mass of corresponding particle. Furthermore, the "review" contains only the sketch of Sallhofer's assertions without any analysis. Moreover, the review in [116] was enough incomplete because not includes the most interesting Maxwell--Dirac additional relationships in the case of nonzero mass! However, at that time the book [3] and the corresponding publications [11--14] were already well-known. Recently, in the form of the article [117], another "review" with similar features was published. It contains the related subjects of longitudinal electromagnetic waves and the generalization of classical Maxwell electrodynamics. The answer is very simple. If somebody cannot understand and learn the classical electrodynamics he starts to generalize this theory.

Thus, it is better to speak carefully that derivation of the Dirac equation in [71--74, 101] and here above is only partially based on mathematics and includes so called physical intuition. Nevertheless, let us note that many important steps in physics were made only on the basis of physical intuition without any appealing to mathematical derivations. And the fine structure of the hydrogen spectrum for the first time was received in [73, 74] in pure classical electrodynamics without the formalism of quantum mechanics.

Finally, the results considered above are very interesting and can be useful for further investigations and investigators. It was the useful step for the mathematically well-defined consideration in next item.

\textbf{34. Origin in maximally symmetrical form of the Maxwell equations.} More then twenty years ago we already presented our own independent derivation of the Dirac equation [106, 107, 113, 114]. The Dirac equation was derived from slightly generalized Maxwell equations with gradient-like current and charge densities. Such Maxwell system includes magnetic gradient-like current and charge densities (in another interpretation system contains additional scalar field). This form of the Maxwell equations, which is directly linked to the Dirac equation, is the maximally symmetrical variant of Maxwell system. Such Maxwell equations are invariant with respect to a 256-dimensional algebra (the well-known algebra of conformal group has only 15 generators). Of course, we derived only massless Dirac equation.

In any case we were not inside the idea of generalization of the standard Maxwell electrodynamics in its ordinary domain of application. Our goal was to adapt the classical electrodynamics in order to described the microword phenomena, such as the fine structure of the hydrogen spectrum and evident consequences. It was well-known in the year of 2002 that the Maxwell classical and quantum electrodynamics needs the adaption to the strong field and to the singularities as the black holes, unexpected phenomena as dark matter and dark energy. In these old times and today we never doubt in classical Maxwell electrodynamics and its natural mapping to general relativity inside the Solar system.

Nevertheless, the principle of correspondence between a generalizations and ordinary classical electrodynamics should be demonstrated in approaches as [108, 111, 112]. Here it is the transition from microworld in [106, 107, 113, 114] to macro phenomena of ordinary classical electrodynamics. This transition causes the loosing symmetry in the Maxwell equations.

Note that J.C. Maxwell derived a system of equations for describing electromagnetic phenomena on the basis of a generalized rewriting of all known electrodynamics laws of Faraday, Ampere, Weber, etc., as well as from the principle of symmetry. Looking for the equations for inneratomic problems in the framework of classical electrodynamics we recall the Maxwell's idea about symmetry principle. By analogy with Maxwell's suggestions in the articles [106, 107, 113, 114], perhaps the most symmetrical form of the Maxwell equations with 256 dimensional invariance algebra is introduced. It is precisely this system of Maxwell equations that is directly related to the massless Dirac equation and can describe the spectrum of hydrogen. The massless Dirac equation follows from such non-ordinary Maxwell equations.

Consider this way of the Dirac equation derivation in some details.

The above mentioned equations for the system of electromagnetic and scalar fields $\overrightarrow{(E},\overrightarrow{H,}%
E^0,H^0)$ have the form:
\begin{equation}
\begin{array}{c}
\partial _0\overrightarrow{E}=\mathrm{curl}\overrightarrow{H}-\mathrm{grad}E^0,\quad
\partial _0\overrightarrow{H}=-\mathrm{curl}\overrightarrow{E}-\mathrm{grad}H^0, \\
\mathrm{div}\overrightarrow{E}=-\partial _0E^0,\quad \mathrm{div}%
\overrightarrow{H}=-\partial _0H^0.
\end{array}
\label{eq24}
\end{equation}
\noindent Here and below the system of units $\hbar =c =1$ is used. The equations (24) are nothing more than the weakly generalized Maxwell equations ($\varepsilon =\mu =1$) with gradient-like electric and magnetic sources $j_\mu ^e=-\partial _\mu E^0,$ $j_\mu ^\mathrm{mag}=-\partial _\mu H^0$, i.e.
\begin{equation}
\overrightarrow{j}_e=-\mathrm{grad}E^0, \, \overrightarrow{j} _\mathrm{mag}=-%
\mathrm{grad}H^0, \quad \rho _e=-\partial _0E^0, \, \rho _\mathrm{mag}=-\partial
_0H^0.
\label{eq25}
\end{equation}

In terms of complex 4-component object
\begin{equation}
\mathcal{E}\equiv \left|
\begin{array}{c}
\overrightarrow{\mathcal{E}} \\
\mathcal{E}^0
\end{array}
\right| =\mathrm{column}\left| E^1-iH^1,E^2-iH^2,E^3-iH^3,E^0-iH^0\right| ,
\label{eq26}
\end{equation}
and, further, in terms of following complex tensor
\begin{equation}
\mathbb{E}=(\mathbb{E}^{\mu \nu })\equiv \left|
\begin{array}{cccc}
0 & \mathcal{E}^1 & \mathcal{E}^2 & \mathcal{E}^3 \\
-\mathcal{E}^1 & 0 & i\mathcal{E}^3 & -i\mathcal{E}^2 \\
-\mathcal{E}^2 & -i\mathcal{E}^3 & 0 & i\mathcal{E}^1 \\
-\mathcal{E}^3 & i\mathcal{E}^2 & -i\mathcal{E}^1 & 0
\end{array}
\right|
\label{eq27}
\end{equation}
equations (24) can be rewritten in the manifestly covariant forms
\begin{equation}
\partial _\mu \mathcal{E}_\nu -\partial _\nu \mathcal{E}_\mu +i\varepsilon
_{\mu \nu \rho \sigma }\partial ^\rho \mathcal{E}^\sigma =0,\quad \partial
_\mu \mathcal{E}^\mu =0,
\label{eq28}
\end{equation}
\noindent (vector form) and tensor-scalar form:
\begin{equation}
\partial _\nu \mathbb{E}^{\mu \nu }=\partial ^\mu \mathcal{E}^0.
\label{eq29}
\end{equation}
It is useful also to consider the following form of equations (24), (28), (29):
\begin{equation}
(i\partial _0-\overrightarrow{S}\cdot \overrightarrow{p})\overrightarrow{%
\mathcal{E}}-i\mathrm{grad}\mathcal{E}^0=0,\quad \partial _\mu \mathcal{E}%
^\mu =0,
\label{eq30}
\end{equation}
\noindent where $\overrightarrow{S}\equiv (S^j)$ are the generators of irreducible
representation D(1) of the group SU(2):
\begin{equation}
S^1=\left|
\begin{array}{ccc}
0 & 0 & 0 \\
0 & 0 & -i \\
0 & i & 0
\end{array}
\right|, \, S^2=\left|
\begin{array}{ccc}
0 & 0 & i \\
0 & 0 & 0 \\
-i & 0 & 0
\end{array}
\right|, \, S^3=\left|
\begin{array}{ccc}
0 & -i & 0 \\
i & 0 & 0 \\
0 & 0 & 0
\end{array}
\right|; \, \overrightarrow{S}^2=1(1+1)I.
\label{eq31}
\end{equation}

The general solution of equations (24), (28)--(30) was found in [111], their symmetry properties were considered in [105], the application (after formulation in specific medium (7)) to the hydrogen spectrum description was given in [108, 112]. The solution was found in the rigged Hilbert space $\mathrm{S}^{4,4}\subset\mathrm{H}^{4,4}\subset\mathrm{S}^{4,4*}$ directly by application of Fourier method. In terms of helicity amplitudes $c^{\bar{\alpha}}(\overrightarrow{k})$ this solution has the form
\begin{equation}
\label{eq32}
\mathcal{E}\left( x\right) =\int \mathrm{d}^3k\sqrt{\frac{2\tilde{\omega} }{\left(
2\pi \right) ^3}}\left\{
\begin{array}{c}
\left[ c^1e_1+c^3\left( e_3+e_4\right) \right] \mathrm{e}^{-ikx}+ \\
\left[ c^{*2}e_1+c^{*4}\left( e_3+e_4\right) \right] \mathrm{e}^{ikx}
\end{array}
\right\} ,\quad \tilde{\omega} \equiv \sqrt{\overrightarrow{k}^2},
\end{equation}
\noindent where $kx\equiv \tilde{\omega}t-\overrightarrow{k}\overrightarrow{x},\, \bar{\alpha}=1,2,3,4,$ and $4$-component basis vectors $e_\alpha $ are taken in the form
\begin{equation}
\label{eq33}
e_1=\left|
\begin{array}{c}
\overrightarrow{e_1} \\
0
\end{array}
\right| ,\quad e_2=\left|
\begin{array}{c}
\overrightarrow{e_2} \\
0
\end{array}
\right| ,\quad e_3=\left|
\begin{array}{c}
\overrightarrow{e_3} \\
0
\end{array}
\right| ,\quad e_4=\left|
\begin{array}{c}
0 \\
1
\end{array}
\right|.
\end{equation}
\noindent Here the $3$-component basis vectors which, without any loss of generality,
can be taken as
\begin{equation}
\label{eq34}
\overrightarrow{e_1}=\frac 1{\tilde{\omega} \sqrt{2\left( k^1k^1+k^2k^2\right) }%
}\left|
\begin{array}{c}
\tilde{\omega} k^2-ik^1k^3 \\
-\tilde{\omega} k^1-ik^2k^3 \\
i\left( k^1k^1+k^2k^2\right)
\end{array}
\right| ,\quad \overrightarrow{e_2}=\overrightarrow{e_1}^{*},\quad
\overrightarrow{e_3}=\frac{\overrightarrow{k}}{\tilde{\omega}} ,
\end{equation}
are the eigenvectors for the quantum-mechanical helicity operator for the
spin $s=1$.

Note that if the quantities $E^0,H^0$ in equations (24) are some given
functions, for which the representation
\begin{equation}
\label{eq35}
E^0-iH^0=\int \mathrm{d}^3k\sqrt{\frac{2\tilde{\omega} }{\left( 2\pi \right) ^3}}%
\left( c^3\mathrm{e}^{-\mathrm{i}kx}+c^4\mathrm{e}^{\mathrm{i}kx}\right)
\end{equation}
\noindent is valid, then (24) are the Maxwell equations with the given sources, $j_\mu ^e=-\partial _\mu E^0,j_\mu ^{mag}=-\partial _\mu H^0$ (namely these 4 currents we call the gradient-like sources). In this case the general solution of the Maxwell equations (24), (28)--(30) with the given sources, as follows from (32), has the form
\begin{equation}
\label{eq36}
\begin{array}{c}
\overrightarrow{E}\left( x\right) =\int \mathrm{d}^3k\sqrt{\frac {\tilde{\omega}}
{2\left( 2\pi \right) ^3}}\left( c^1\overrightarrow{e}_1+c^2\overrightarrow{e%
}_2+\alpha \overrightarrow{e}_3\right) \mathrm{e}^{-\mathrm{i}kx}+c.c, \\
\overrightarrow{H}\left( x\right) =i\int \mathrm{d}^3k\sqrt{\frac {\tilde{\omega}}
{2\left( 2\pi \right) ^3}}\left( c^1\overrightarrow{e}_1-c^2\overrightarrow{e%
}_2+\beta \overrightarrow{e}_3\right) \mathrm{e}^{-\mathrm{i}kx}+c.c,
\end{array}
\end{equation}
\noindent where the amplitudes of longitudinal waves $\overrightarrow{e}_3\exp \left( -%
\mathrm{i}kx\right) $ are $\alpha =c^3+c^4,$ $\beta =c^3-c^4$ and $c^3,c^4$
are determined by the functions $E^0,H^0$ according to the formula (35).

\textit{The following assertion is valid}.

Equations (24), (28)--(30) are directly related to the free massless Dirac equation
\begin{equation}
\label{eq37}
i\gamma ^\mu \partial _\mu \psi (x)=0.
\end{equation}

\textit{The proof is as follows}.

Note at first that there is no reason to appeal here to the stationary case as it was done in
[71--74], where the case with nonzero interaction and mass $m\neq 0$ was considered. Indeed. the
substitution of
\begin{equation}
\label{eq38}
\psi =\left|
\begin{array}{c}
E^3+iH^0 \\
E^1+iE^2 \\
iH^3+E^0 \\
-H^2+iH^1
\end{array}
\right| =U\mathcal{E},\, U=\left|
\begin{array}{cccc}
0 & 0 & C_{+} & C_{-} \\
C_{+} & iC_{+} & 0 & 0 \\
0 & 0 & C_{-} & C_{+} \\
C_{-} & iC_{-} & 0 & 0
\end{array}
\right|; \, C_{\mp }\equiv \frac 12(C\mp 1),
\end{equation}
\noindent ($C$ is the operator of complex conjugation, $C\mathcal{E}^{1}=\mathcal{E}^{1*}$; complex field strength $\mathcal{E}$ is known from (26)) into Dirac equation (37) with $\gamma $ matrices in standard Pauli--Dirac representation (3) guarantees its transformation into the generalized Maxwell equations (24), (28)--(30).

Moreover, the substitution of column (26)
\begin{equation}
\label{eq39}
\mathcal{E} =\left|
\begin{array}{c}
E^1-iH^1 \\
E^2-iH^2 \\
E^3-iH^3 \\
E^0-iH^0
\end{array}
\right| =U^{-1}\psi,\quad U^{-1}=U^{\dagger }=\left|
\begin{array}{cccc}
0 & C_{+} & 0 & C_{-} \\
0 & iC_{-} & 0 & iC_{+} \\
C_{+} & 0 & C_{-} & 0 \\
C_{-} & 0 & C_{+} & 0
\end{array}
\right|,
\end{equation}
into the equation
\begin{equation}
\label{eq40}
\widetilde{\gamma }^\mu \partial _\mu
\mathcal{E}(\textit{x})=0
\end{equation}
with $\widetilde{\gamma }^\mu = U^{-1}\gamma^{\mu}U$ matrices
\begin{eqnarray}
\label{eq41}
\widetilde{\gamma }^0 &=&\left|
\begin{array}{cccc}
1 & 0 & 0 & 0 \\
0 & 1 & 0 & 0 \\
0 & 0 & 1 & 0 \\
0 & 0 & 0 & -1
\end{array}
\right| C,\quad \widetilde{\gamma }^1=\left|
\begin{array}{cccc}
0 & 0 & 0 & 1 \\
0 & 0 & -i & 0 \\
0 & i & 0 & 0 \\
-1 & 0 & 0 & 0
\end{array}
\right| C, \\
\widetilde{\gamma }^2 &=&\left|
\begin{array}{cccc}
0 & 0 & i & 0 \\
0 & 0 & 0 & 1 \\
-i & 0 & 0 & 0 \\
0 & -1 & 0 & 0
\end{array}
\right| C,\quad \widetilde{\gamma }^3=\left|
\begin{array}{cccc}
0 & -i & 0 & 0 \\
i & 0 & 0 & 0 \\
0 & 0 & 0 & 1 \\
0 & 0 & -1 & 0
\end{array}
\right| C,  \nonumber
\end{eqnarray}
guarantees its transformation into the generalized Maxwell equations (24), (28)--(30) as well.

Proved assertions demonstrate both the derivation of the slightly generalized Maxwell equations from the massless Dirac equation and the derivation of the massless Dirac equation from such kind of Maxwell system.

Note specially that ordinary Maxwell equation follow from (24), (28)--(30), (40) in the case $E^{0}=H^{0}=0$.

The transitions (38), (39) are based on unitary transformation. The unitary properties of the operator \textit{U} can be verified easily by taking into account that the equations
\begin{equation}
\label{eq42}
\left( AC\right) ^{\dagger }=CA^{\dagger },\quad aC=Ca^{*},\quad \left(
aC\right) ^{*}=Ca
\end{equation}
hold for an arbitrary matrix $A$ and a complex number $a$. We note that in
the real algebra (i. e. the algebra over the field of real numbers) and in
the Hilbert space of quantum mechanical amplitudes the operator $U$ in (38), (39) has all
properties of unitarity: $UU^{-1}=U^{-1}U=1,$ $U^{-1}=U^{\dagger }$, plus
linearity.

Due to the unitarity of the operator $U$ the $\widetilde{\gamma }%
^\mu $ matrices (41) still obey the anticommutation relations of Clifford--Dirac algebra
\begin{equation}
\label{eq43}
\widetilde{\gamma }^\mu \widetilde{\gamma }^\nu +\widetilde{ \gamma }^\nu
\widetilde{\gamma }^\mu =2g^{\mu \nu }
\end{equation}
and have the same Hermitian properties as the Pauli - Dirac $\gamma ^\mu $ matrices (3):
\begin{equation}
\label{eq44}
\widetilde{\gamma }^{0\dagger }=\widetilde{\gamma }^0,\quad \widetilde{%
\gamma }^{k\dagger }=-\widetilde{\gamma }^k.
\end{equation}

The complete set of 8 transformations like (38), which relates generalized Maxwell equations (24) and massless Dirac equation (37), was found in [102, 106]. The explicit form is as follows
$$\psi^\mathrm{I} =\left|
\begin{array}{c}
E^3+iH^0 \\
E^1+iE^2 \\
iH^3+E^0 \\
-H^2+iH^1
\end{array}
\right|, \,
\psi^\mathrm{II} =\left|
\begin{array}{c}
iE^3-H^0 \\
iE^1-E^2 \\
-H^3+iE^0 \\
-H^1-iH^2
\end{array}
\right|, \,
\psi^\mathrm{III} =\left|
\begin{array}{c}
iE^1+E^2 \\
-iE^3-H^0 \\
-H^1+iH^2 \\
H^3+iE^0
\end{array}
\right|,$$
\begin{equation}
\label{eq45}
\psi^\mathrm{IV} =\left|
\begin{array}{c}
-E^1+iE^2 \\
E^3-iH^0 \\
-iH^1-H^2 \\
iH^3-E^0 \\
\end{array}
\right|, \,
\psi^\mathrm{V} =\left|
\begin{array}{c}
-H^3+iE^0 \\
-H^1-iH^2 \\
iE^3-H^0 \\
iE^1-E^2 \\
\end{array}
\right|, \,
\psi^\mathrm{VI} =\left|
\begin{array}{c}
-H^1+iH^2 \\
H^3+iE^0 \\
iE^1+E^2 \\
-iE^3-H^0 \\
\end{array}
\right|,
\end{equation}
$$\psi^\mathrm{VII} =\left|
\begin{array}{c}
iH^3+E^0 \\
-H^2+iH^1 \\
E^3+iH^0 \\
E^1+iE^2 \\
\end{array}
\right|, \,
\psi^\mathrm{VIII} =\left|
\begin{array}{c}
-iH^1-H^2 \\
iH^3-E^0 \\
-E^1+iE^2 \\
E^3-iH^0 \\
\end{array}
\right|.$$
Relationship between the generalized Maxwell equations (24), (28)--(30)) and massless Dirac equation (37) is considered here in the way similar to the [109] and [105]. The \textit{proof} that the number of such transformation is eight is similar to the proof given above for the set (20). This number is eight, because the number of pure matrix Pauli--G$\ddot{\mathrm{u}}$rsey--Ibragimov operators [102--104] $ \{\gamma^{2}\hat{C}, \, i\gamma^{2}\hat{C}, \,
\gamma^{2}\gamma^{4}\hat{C}, \, i\gamma^{2}\gamma^{4}\hat{C}, \,
\gamma^{4},  \, i\gamma^{4}, \, i, \, \mathrm{I} \}$, which leave the massless Dirac equation being invariant, is eight.

Note that equations (24), (28)--(30), (40) are the maximally symmetrical form of the Maxwell system. Indeed, here both Dirac-like and Maxwell-like symmetries are natural and are ready for application. The Bose symmetries of the massless Dirac equation and Fermi symmetries of the the slightly generalized Maxwell equations together with other corresponded results were considered in [105--107].

Further, contrary to the formalism of the Item 31 above, here there is one-to-one correspondence between the solutions of the massless Dirac equation (37) and the slightly generalized Maxwell equations (24), (28)--(30). Therefore, the presented here derivation of the massless Dirac equation is direct, simple and well-defined mathematically.

\textbf{35. Derivation of the Dirac equation from the relativistic canonical quantum mechanics.} In our recent papers [15, 16, 56--58] and [3] the Dirac equation with nonzero mass has been derived from the quantum-mechanical relativistic equation for the fermion-antifermion particle doublet of spins $s=1/2$. We proposed to call such 4-component relativistic equation, which is the starting point of this derivation, as the Schr$\mathrm{\ddot{o}}$dinger--Foldy equation (the one-component case is often called as the spinless Salpeter equation [38--54] or as the equation for L$\mathrm{\acute{e}}$vy flight [55]). Corresponded model of the physical reality, which is based on such equation of motion, was called by us as relativistic canonical quantum mechanics (RCQM) of particle-antiparticle doublet. The arbitrary spins and arbitrary dimensions have been considered.

For the partial case of spin 1/2 fermion-antifermion doublet the 4-component Sch\"odinger--Foldy equation is given by
\begin{equation}
\label{eq46}
i\partial_{t}f(x)=\sqrt{m^{2}-\Delta}f(x), \quad f=\left|
{{\begin{array}{*{20}c}
 f^{1} \hfill  \\
 f^{2} \hfill  \\
 f^{3} \hfill  \\
 f^{4} \hfill  \\
\end{array} }} \right|,
\end{equation}
\noindent and is considered in the rigged Hilbert space $\mathrm{S}^{3,4}\subset\mathrm{H}^{3,4}\subset\mathrm{S}^{3,4*}$, where $\mathrm{S}^{3,4}$ is the 4-component Schwartz test function space over the space
$\mathrm{R}^{3}\subset \mathrm{M}(1,3)$ and
$\mathrm{H}^{3,4}$ is the Hilbert space of the
4-component square-integrable functions over the
$x\in\mathrm{R}^{3}\subset \mathrm{M}(1,3)$
\begin{equation}
\label{eq47}
\mathrm{H}^{3,4}=\mathrm{L}_{2}(\mathrm{R}^3)\otimes\mathrm{C}^{\otimes 4}=\{f=(f^{4}):\mathrm{R}^{3}\rightarrow\mathrm{C}^{\otimes 4}; \, \int d^{3}x|f(t,\overrightarrow{x})|^{2} <\infty\},
\end{equation}
\noindent where $d^{3}x$ is the Lebesgue measure in the space $\mathrm{R}^{3}\subset \mathrm{M}(1,3)$ of the eigenvalues of the position operator $\overrightarrow{x}$ of the Cartesian coordinate of the particle in an arbitrary-fixed inertial frame of references, $\mathrm{M}(1,3)$ is the Minkowski space. Further, $\mathrm{S}^{3,4*}$ is the space of the
4-component Schwartz generalized functions. The space $\mathrm{S}^{3,4*}$ is conjugated to that of the
Schwartz test functions $\mathrm{S}^{3,4}$ by the corresponding topology (see, e. g., [118]).

The following assertion is valid.

The link between the quantum-mechanical equation (46) and the Dirac equation
\begin{equation}
\label{eq48}
i\partial _0 \psi(x)=(\overrightarrow{\alpha}\cdot\overrightarrow{p}+\beta m)\psi(x), \quad \overrightarrow{\alpha}\equiv \gamma^{0}\overrightarrow{\gamma}, \, \beta \equiv \gamma^{0},
\end{equation}
\noindent is given by the operator
\begin{equation}
\label{eq49}
V=\frac{i\gamma^{\ell}\partial_{\ell}+\widehat{\omega}+m}{\sqrt{2\widehat{\omega}(\widehat{\omega}+m)}}\left|
\begin{array}{cccc}
1 & 0 & 0 & 0 \\
0 & 1 & 0 & 0 \\
0 & 0 & C & 0 \\
0 & 0 & 0 & C \\
\end{array}
\right|, \quad
V^{-1}=\left|
\begin{array}{cccc}
1 & 0 & 0 & 0 \\
0 & 1 & 0 & 0 \\
0 & 0 & C & 0 \\
0 & 0 & 0 & C \\
\end{array}
\right|\frac{-i\gamma^{\ell}\partial_{\ell}+\widehat{\omega}+m}{\sqrt{2\widehat{\omega}(\widehat{\omega}+m)}},
\end{equation}
$$VV^{-1}=V^{-1}V=\mathrm{I}, \quad \widehat{\omega}=\sqrt{-\Delta+m^{2}},$$
\noindent where $C$ is the operator of complex conjugation, $C\psi^{1}=\psi^{1*}$, (the operator of involution in $\mathrm{H}^{3,1}$). The following relationships are valid
\begin{equation}
\label{eq50}
V(\partial_{0}+i\widehat{\omega})V^{-1}=\partial _0 +i(\overrightarrow{\alpha}\cdot\overrightarrow{p}+\beta m),
\end{equation}
\noindent for the anti-Hermitian operators of corresponding equations, and
\begin{equation}
\label{eq451}
\psi^{\mathrm{Dirac}}(x)=Vf^{\mathrm{Sr-Foldy}}(x),
\end{equation}
\noindent for the corresponding general solutions. The vice-verse relations are valid as well, but for the derivation of the Dirac equations directly the formulas (50), (51) are used.

The proof of this assertion is fulfilled by direct calculations of (50) and (51). In order to verify (51) the explicit forms of corresponding general solutions are used. These solutions are given below.

Note that due to the presence of the operator $C$ our extended FW operator (49) is well-defined only in application to anti-Hermitian operators. For the possibility and mathematical correctness in using the anti-Hermitian operators in quantum theory see, e.g., [119, 120].

In (51) the general solution $f^{\mathrm{Sr-Foldy}}(x)$ of the Schr$\mathrm{\ddot{o}}$dinger--Foldy equation of motion (46) has the form
\begin{equation}
\label{52}
f^{\mathrm{Sr-Foldy}}(x)= \left|
{{\begin{array}{*{20}c}
 f_{\mathrm{part}} \hfill  \\
 f_{\mathrm{antipart}} \hfill  \\
\end{array} }} \right| =\frac{1}{\left(2\pi\right)^{\frac{3}{2}}}\int d^{3}k e^{-ikx}
\end{equation}
$$
[a^{1}(\overrightarrow{k})\mathrm{d}_{1}+a^{2}(\overrightarrow{k})\mathrm{d}_{2}+a^{3}(\overrightarrow{k})\mathrm{d}_{3}+a^{4}(\overrightarrow{k})\mathrm{d}_{4}],
$$
where the orts $\left\{\mathrm{d}_{\alpha}\right\}$ of the Cartesian basis are given by
\begin{equation}
\label{eq53}
\mathrm{d}_{1} = \left|
\begin{array}{cccc}
 1 \\
 0 \\
 0 \\
 0 \\
\end{array} \right|, \, \mathrm{d}_{2} = \left|
\begin{array}{cccc}
 0 \\
 1 \\
 0 \\
 0 \\
\end{array} \right|, \,
\mathrm{d}_{3} = \left|
\begin{array}{cccc}
 0 \\
 0 \\
 1 \\
 0 \\
\end{array} \right|, \,
\mathrm{d}_{4} = \left|
\begin{array}{cccc}
 0 \\
 0 \\
 0 \\
 1 \\
\end{array} \right|,
\end{equation}
and the following notations
\begin{equation}
\label{eq54}
kx\equiv \widetilde{\omega} t -\overrightarrow{k}\overrightarrow{x}, \quad \widetilde{\omega} \equiv \sqrt{\overrightarrow{k}^{2}+m^{2}},
\end{equation}
\noindent are used.

The interpretation of the amplitudes in the general solution (52) follows from equations
\begin{equation}
\label{eq55} \widehat{\overrightarrow{p}}e^{-ikx}\mathrm{d}_{\bar{\alpha}} =
\overrightarrow{k}e^{-ikx}\mathrm{d}_{\bar{\alpha}}, \quad \bar{\alpha} =1,2,3,4,
\end{equation}
\begin{equation}
\label{eq56}
s^{3}\mathrm{d}_{1} = \frac{1}{2}\mathrm{d}_{1}, \, s^{3}\mathrm{d}_{2} = -\frac{1}{2}\mathrm{d}_{2}, \, s^{3}\mathrm{d}_{3} = -\frac{1}{2} \mathrm{d}_{3}, \, s^{3}\mathrm{d}_{4} = \frac{1}{2}\mathrm{d}_{4},
\end{equation}
\begin{equation}
\label{eq57}
g\mathrm{d}_{1} = -e\mathrm{d}_{1}, \,
g\mathrm{d}_{2} = -e\mathrm{d}_{2}, \,  g\mathrm{d}_{3} =
e\mathrm{d}_{3}, \, g\mathrm{d}_{4} = e\mathrm{d}_{4},
\end{equation}
on eigen vectors and eigenvalues of the operators of stationary complete set $(\widehat{\overrightarrow{p}}, \, s_{z}\equiv s^{3}, \, g)$. Thus, the functions $a^{1}(\overrightarrow{k}), \, a^{2}(\overrightarrow{k})$ are the momentum-spin amplitudes of the particle (e. g., electron) with the momentum $\widehat{\overrightarrow{p}}$, sign of the charge ($-e$) and spin projections ($\frac{1}{2}, \, -\frac{1}{2}$), respectively. Further, the functions $a^{3}(\overrightarrow{k}), \, a^{4}(\overrightarrow{k})$ in (52) are the momentum-spin amplitudes of the antiparticle (e. g., positron) with the momentum $\widehat{\overrightarrow{p}}$, sign of the charge ($+e$) and spin projections ($-\frac{1}{2}, \, \frac{1}{2}$), respectively.

Taking into account the Pauli principle and the fact that experimentally positron is observed as the mirror reflection of an electron, the operators of the charge sign and the spin of the s=(1/2,1/2) particle-antiparticle doublet are taken in the form
\begin{equation}
\label{eq58} g\equiv-\gamma^0 = \left| {{\begin{array}{*{20}c}
 -\mathrm{I}_{2} \hfill & 0 \hfill \\
 0 \hfill & \mathrm{I}_{2} \hfill \\
\end{array} }} \right|, \quad \overrightarrow{s} =
\frac{1}{2}\left| {{\begin{array}{*{20}c}
 \overrightarrow{\sigma} \hfill  0 \hfill \\
 0 -C\hfill\overrightarrow{\sigma}\hfill C \\
\end{array} }} \right|,
\end{equation}
\noindent where $\overrightarrow{\sigma}$ are the standard Pauli matrices (1), $C$ is the operator of complex conjugation, $\mathrm{I}_{2}$ is $2\times2$ unit matrix.

In the choice of the spin (58) the principle of correspondence and heredity with the FW representation is used, where the particle-antiparticle doublet spin operator is given by
\begin{equation}
\label{eq59} \overrightarrow{s}_{\mathrm{FW}} =
\frac{1}{2}\left| {{\begin{array}{*{20}c}
 \overrightarrow{\sigma} \, \hfill   0  \hfill \\
 0  \, \hfill \overrightarrow{\sigma} \hfill  \\
\end{array} }} \right|.
\end{equation}
\noindent The link between the spins (58) and (59) $v \overrightarrow{s}_{\mathrm{FW}}v=\overrightarrow{s}, \, v\overrightarrow{s}v= \overrightarrow{s}_{\mathrm{FW}}$ is given by the transformation operator $v$
\begin{equation}
\label{eq60}
v=v^{-1}=\left|
\begin{array}{cccc}
1 & 0 & 0 & 0 \\
0 & 1 & 0 & 0 \\
0 & 0 & C & 0 \\
0 & 0 & 0 & C \\
\end{array}
\right|, \quad vv^{-1}=v^{-1}v=\mathrm{I},
\end{equation}
\noindent which extends the FW transformation in (49). Note that transformation $v$ is valid only for the case of anti-Hermitian form of spins (58), (59).

In (51) the general solution $\psi^{\mathrm{Dirac}}(x)$ of the Dirac equation (48) is given by
\begin{equation}
\label{eq61}
\psi(x)=V^{+}\phi(x)= \frac{1}{\left(2\pi\right)^{\frac{3}{2}}}\int d^{3}k\left[e^{-ikx}a^{\mathrm{r}}(\overrightarrow{k})\mathrm{v}^{-}_{\mathrm{r}}(\overrightarrow{k})+e^{ikx}a^{*\mathrm{\check{r}}}(\overrightarrow{k})\mathrm{v}^{+}_{\mathrm{\check{r}}}(\overrightarrow{k})\right],
\end{equation}
$$\mathrm{r}=(1,2), \, \mathrm{\check{r}}=(3,4).$$
\noindent The amplitudes $a^{\alpha}(\overrightarrow{k})$ in (61) are the same as in (52) in the corresponding RCQM. The basis vectors are changed and now have the form
\begin{equation}
\label{eq62} \mathrm{v}^{-}_{1}(\overrightarrow{k}) = N\left|
\begin{array}{cccc}
 \widetilde{\omega}+m \\
 0 \\
 k^{3} \\
 k^{1}+ik^{2} \\
\end{array} \right|, \quad
\mathrm{v}^{-}_{2}(\overrightarrow{k}) = N\left|
\begin{array}{cccc}
 0 \\
 \widetilde{\omega}+m \\
 k^{1}-ik^{2} \\
 -k^{3} \\
\end{array} \right|,
\end{equation}
$$\mathrm{v}^{+}_{3}(\overrightarrow{k}) = N\left|
\begin{array}{cccc}
 k^{3} \\
 k^{1}+ik^{2} \\
 \widetilde{\omega}+m \\
 0 \\
\end{array} \right|, \quad
\mathrm{v}^{+}_{4}(\overrightarrow{k}) = N\left|
\begin{array}{cccc}
  k^{1}-ik^{2} \\
 -k^{3} \\
  0 \\
 \widetilde{\omega}+m \\
\end{array} \right|,$$
\noindent where
\begin{equation}
\label{eq63} N\equiv
\frac{1}{\sqrt{2\widetilde{\omega}(\widetilde{\omega}+m)}}, \quad
\widetilde{\omega}\equiv \sqrt{\overrightarrow{k}^{2}+m^{2}}.
\end{equation}
\noindent Thus, the orts
$\mathrm{v}^{\pm}_{\alpha}(\overrightarrow{k})$ are the standard
4-component Dirac spinors. Their conditions of orthonormalization
and completeness are well known, see, e. g. [62].

Hence, the Dirac equation (48) is derived from the quantum-mechanical equation (46).

Note that this link is working in both sides. Similarly the equation (40) can be derived from the Dirac equation (48). Therefore, in this sense the equation (46) is the quantum-mechanical limit of the field-theoretical Dirac equation.

\begin{center}
\textbf{Derivation of the Dirac-like equations}
\end{center}

\textbf{36. On the derivation of the Dirac equation in N and six spatial dimensions.} The start of the Dirac equation derivation in $N$ spatial dimensions was given in [59] in the procedure of derivation of the ordinary 4-component Dirac equation. Trying to explain for students the principles of quantum mechanics authors suggested the contemporary point of view for the Dirac derivation [1] of his own equation. Automatically they had derived the Dirac equation for $N$-component vector and with $N\times N$ matrices. The ordinary 4-component Dirac equation follows from such formalism in the case of a minimal size $4\times 4$ of corresponding alpha matrices.
Indeed, the first appearance of the Dirac equation in [59] has the form (1.14) from [59]:
\begin{equation}
\label{eq64}
i\frac{\partial\psi_{\sigma}}{\partial t}=-i\sum_{\tau=1}^{N}(\alpha_{1}\frac{\partial}{\partial x_{1}}+\alpha_{2}\frac{\partial}{\partial x_{2}}+\alpha_{3}\frac{\partial}{\partial x_{3}})_{\sigma\tau}\psi_{\tau}+\sum_{\tau=1}^{N}\beta_{\sigma\tau}m\psi_{\tau}
\end{equation}
$$
=\sum_{\tau=1}^{N}H_{\sigma\tau}\psi_{\tau},
$$
where
\begin{equation}
\label{eq65}
\psi = N\left|
\begin{array}{cccc}
 \psi_{1} \\
 \cdot \\
 \cdot \\
 \cdot \\
 \psi_{N} \\
\end{array} \right|,
\end{equation}
and the constant coefficients $\alpha_{r}, \beta$ are $N\times N$ matrices.

Today, taking into account the theory of the $N$-dimensional gamma matrix representations of the Clifford algebras [121--126] the $N$-component Dirac equation can be constructed in the N-dimensional space-time. Such minimal generalization of the derivation in [1] was considered in [60, 127, 128] and was called the Dirac equation in $D+1$ space-time or the Dirac equation in $D$ spatial dimensions
\begin{equation}
\label{eq66}
\sum_{\mu=0}^{N}i\gamma^{\mu}(\partial_{\mu}+ieA_{\mu})\psi(\vec{x},t)=m\psi(\vec{x},t),
\end{equation}
where m is the mass of the particle, $N\times N$ matrices $\gamma^{\mu}$ satisfies the anticommutative relations $\gamma^{\mu}\gamma^{\nu}+\gamma^{\nu}\gamma^{\mu}=2g^{\mu\nu}$ of Clifford algebra representation, $g^{\mu\nu}$ is $N\times N$ metric tensor, $g^{\mu\nu}=\delta_{\mu\nu}(\mu=0)$ or $g^{\mu\nu}=-\delta_{\mu\nu}(\mu\neq 0)$. In the papers [127--129] the partial case of potentials $A_{\mu}$ in equation (1) was considered, where $eA_{0}=V(r)$ and $eA_{\mu}=0 (\mu\neq 0)$. The separation of variables in hyperspherical coordinates of the real $D$-dimensional space was fulfilled and the cooresponding radial equations were solved. The free noninteracting Dirac equation (66) was not under consideration.

We started the detailed investigation of the free Dirac equation in the spaces of higher dimensions in the paper [129], where the generalized Dirac equation related to 7-component space-time with one time coordinate and six space coordinates has been introduced and solved. Three 8-component Dirac equations from one and the same 64-dimensional gamma matrix representations of the Clifford algebra have been derived. It was demonstrated that each individual case of space dimension has its own specifics and the formal generalization of the Dirac equation for $N$-dimensional space-time is not perfect.

The 8-component Dirac equation in 7-component space-time was derived in the paper [129] on the basis of the method developed and applied in our articles [3, 15, 16, 56] and already described here above as derivation of the Dirac equation from the RCQM.

\textbf{37. Derivation of the fractional Dirac equation.} The fractional Dirac equation of order 2/3 was introduced for the first time in [130, 131] in the form as follows
\begin{equation}
\label{eq67}
-\gamma^{\alpha}\mathrm{D}_{\alpha}^{2/3}\Psi(x)=(m)^{2/3}\Psi(x),
\end{equation}
where $\mathrm{D}_{\alpha}^{2/3}$ are the fractional derivatives [132] of order 2/3, the (complex) matrices $\gamma^{\alpha}$ are $N\times N$,  $\Psi(x)$ is a complex $N$-spinor and $9\times 9$ is the smallest dimension of gamma matrices (see [130] for details). The explicit form of the unitary choice of four $9\times 9$ gamma matrices was given in [131].

The fractional Dirac equation was derived in [130] as a "cube root" of Klein--Gordon equation (operator $-\gamma^{\alpha}\mathrm{D}_{\alpha}^{2/3}$ is a "cube root" of $P^{2}$).

Raspini's formulation [130, 131] was extended to derive fractional Dirac equations of order $m/n$, where $m\leq n$ are two arbitrary integers.

Moreover, the fractional Dirac equation was derived in the paper [133] using a fractional variational principle and the fractional Euler--Lagrange equations of motion.

Nevertheless, the fractional Dirac equation never be such useful as the ordinary Dirac equation. The fractional Dirac equation is essentially nonlinear integro-differential equation. The fractional Schr$\mathrm{\ddot{o}}$dinger equation has some better perspectives.

\textbf{38. On the derivation of the Dirac-like equations for spin 3/2, 2 and arbitrary spin.} The Dirac-like equations without redundant components for the particle having spin 3/2, 2 and arbitrary spin were derived in our papers [15, 16, 58, 134]. The results were reviewed and systemized in the monograph [3]. The derivation was fulfilled as the generalization of the method considered here above in Item 33. The start of such derivation for an arbitrary spin from the $N$-component variant of the Schr$\mathrm{\ddot{o}}$dinger--Foldy equation (46) has been given, see, e.g., [3].

\section{Conclusions and perspectives}

Original investigations of such important problem of modern theoretical physics as the nature and essence of the Dirac equation are presented. A variety of approaches and appeals to independent mathematical formalisms are demonstrated. In particular, the derivation of the Dirac equation from the principles of information processing, or the way to this equation on the basis of conformal differential geometry and our own derivations (in general 38 different methods), are considered.

At first consider briefly the derivation of the Dirac equation given in Item 35.

Note that the RCQM, which is based on Shr\"odinger--Foldy equation (46) and its general solution (52), is free from difficulties of the Dirac model and, moreover, is free from the difficulties of the FW model (contradictions of the canonical FW representation with main principles of quantum mechanics are considered briefly in Item 9). Here in the solution (52) the negative energies are absent and the advantages of the spin operator (58) in comparison with the FW spin (59) are evident as well. As a consequences in [3] on the basis of (46) the self-consistent relativistic quantum-mechanical model has been constructed, which is similar to the von Neumann's non-relativistic consideration [135].

Furthermore, in (49) the natural generalization of the FW transformation is considered. The extended FW operator, which relates the Dirac and pure quantum-mechanical models, is considered.

Let us mention that the Shr\"odinger--Foldy equation (46) is the main equation of the RCQM of spin s=(1/2,1/2) particle-antiparticle doublet. Hence, the Dirac equation has been derived from the more fundamental model of the same physical reality, which is formulated as the RCQM of the spin s=(1/2,1/2) particle-antiparticle doublet.

In general, the above demonstrated many features of the Dirac equation prove the continuous interest to the problem.

The performed analysis gives an opportunity to compare different methods of the Dirac equation derivation. Probably the simplest is considered
in [62] approach (Item 3), which is based on the start from the Klein--Gordon equation. Nevertheless, such approach becomes quite difficult, if to begin from the Klein-Gordon equation derivation. It is more interesting to compare the fundamentals of different conclusions, their physical and philosophical significance. In this case the Item 35 should be mentioned, where the Dirac equation is derived from the RCQM of the fermion-antifermion doublet, which is more natural model of physical reality than the Dirac model of the spinor field at the classical level. However, the most fundamental is the derivation of the Dirac equation from the principle of the Euler--Lagrange least action in the application to the classical (not quantum) spinor field (Item 4).

A review of the 38 different derivations of the Dirac equation demonstrates (in particular) that presented in [15, 16, 56--58, 134] and in the book [3] method of the Dirac equation (and the Dirac-like equation for arbitrary spin) derivation is original and new. In general, the presented review demonstrates evidently that the Dirac equation is related to the fundamental laws and principles of physics.

It follows from the above given consideration that suggestions of new methods of the Dirac equation derivation are not stopped and \textit{to be continued}.

\textbf{A criterion for the usefulness of one or another derivation of the Dirac equation.}  \textit{The derivation should be independent from the Dirac theory.} Many "modern methods" of the Dirac equation derivation are impossible in principle without the preliminary knowledge of his equation, corresponding Dirac's suggestions, new objects introduced by him and spinor field formalism in general. Indeed, if somebody appeals to such objects as Dirac spinors, one or another conservation law for the Dirac field, Clifford--Dirac algebra, etc., so this kind of derivation is not independent. Such thoughts are possible only after introduction of the Dirac equation and its existence. So what is derived in this case? Our remark is related to many investigators of the Dirac equation derivation, even to considered here in the review. We ask authors to find such methodological and logical kinds of their errors themselves.

\textbf{Suggestions.} Here once more, as in our previous publications [3], [15, 16], [56--58], we suggest to call the main equation of the contemporary $N$-component RCQM as the Schr$\mathrm{\ddot{o}}$dinger--Foldy equation
\begin{equation}
\label{eq68}
i\partial_{t}f(x)=\sqrt{m^{2}-\Delta}f(x), \quad f=\left|
{{\begin{array}{*{20}c}
 f^{1} \hfill  \\
 f^{2} \hfill  \\
 \cdot  \\
 \cdot  \\
 \cdot  \\
 f^{N} \hfill  \\
\end{array} }} \right|.
\end{equation}
Equation (68) is considered in the rigged Hilbert space $\mathrm{S}^{3,N}\subset\mathrm{H}^{3,N}\subset\mathrm{S}^{3,N*}$.
Indeed, in the Items 1 and 35 the significance, usefulness and different applications of this equation were demonstrated.

Furthermore, near ten years ago in the review article [136] a suggestion to call the object $\overrightarrow{\mathcal{E}}=\mathrm{column}\left| E^1-iH^1,E^2-iH^2,E^3-iH^3 \right|$ as the Riemann--Silberstein vector has been given. Note the existence and significance of much more powerful covariant 4-vector (26) considered not only here above in Item 34 but in many publications, see, e.g. our papers [111, 113, 114] and [105--107].

For the first time the corresponding link between the Dirac and the Maxwell equations has been introduced in the paper [137]. Indeed, C.G. Darwin was the first who suggested two of the eight substitutions (45) linking the Dirac and the Maxwell equations. The contemporary form (26) was suggested in [111]. Taking into account above given facts we suggest here to call the vector [26] as the Darwin vector!

\textbf{The main conclusion} from the above considered variety of the approaches to the Dirac equation derivation is the evident signal that the similar variety in the approaches to the modern quantum field theory and standard model is expected as well. Problems of dark matter and dark energy inspired such investigations. For example, new approaches to quantum electrodynamics are the content of the books [138, 3] and the paper [139]. One of the goals of the consideration in the monograph [3] was to present few new approaches of such kind.

\end{document}